\documentclass[namedreferences]{solarphysics}
\usepackage[hyperref,optionalrh,solaromanenum]{spr-sola-addons} 
\usepackage{graphicx}                    
\usepackage{amssymb}                    
\usepackage{color}                       
\usepackage{breakurl}                         
\usepackage{natbib}
\usepackage[breaklinks]{hyperref}
\usepackage{color}
\usepackage[misc]{ifsym}

\newcommand{\MgII}{Mg~{\sc ii}}
\newcommand{\eg}{\textit{e}.\textit{g}.}
\newcommand{\ie}{\textit{i}.\textit{e}.}
\def\arcsec{$^{\prime\prime}$}

\newcommand{\aap}{    {\it Astron. Astrophys.}}

\newcommand{\apj}{    {\it Astrophys. J.}}
\newcommand{\apjs}{    {\it Astrophys. J. Suppl.}}
\newcommand{\apjl}{   {\it Astrophys. J. Lett.}}

\newcommand{\pasj}{   {\it Pub. Astron. Soc. Japan}}

\newcommand{\solphys}{{\it Solar Phys.}}

\begin{document}

\begin{article}

\begin{opening}

\title{Mg~{\sc ii} Lines Observed during the X-class Flare on 29 March 2014 by the {\it Interface Region Imaging Spectrograph}}

%
\author{W.~\surname{Liu}$^{1}$\sep
        P.~\surname{Heinzel}$^{1}$\sep
        L.~\surname{Kleint}$^{2}$\sep
        J.~\surname{Ka\v{s}parov\'{a}}$^{1}$
       }

%
\runningauthor{}
\runningtitle{}

%
  \institute{\Letter\ W.~\surname{Liu}\\
             \quad\ \href{mailto:wenjuan.liu@asu.cas.cz}{wenjuan.liu@asu.cas.cz}\\  
             \vspace{1em}
             \quad\ P.~\surname{Heinzel}\\
             \quad\ \href{mailto:petr.heinzel@asu.cas.cz}{petr.heinzel@asu.cas.cz}\\  
             \vspace{1em}
             \quad\ L.~\surname{Kleint}\\
             \quad\ \href{mailto:lucia.kleint@fhnw.ch}{lucia.kleint@fhnw.ch}\\  
             \vspace{1em}
             \quad\ J.~\surname{Ka\v{s}parov\'{a}}\\
             \quad\ \href{mailto:jana.kasparova@asu.cas.cz}{jana.kasparova@asu.cas.cz}\\  
             \vspace{1em}
             $^{1}$ Astronomical Institute, the Czech Academy of Sciences, 25165 Ond\v{r}ejov, Czech Republic\\
            \vspace{1em}
             $^{2}$ University of Applied Sciences and Arts Northwestern Switzerland, 5210 Windisch, Switzerland\\
             }

\begin{abstract}
\MgII\ lines represent one of the strongest emissions from the chromospheric
plasma during solar flares. In this article, we studied the \MgII\ lines observed 
during the X1 flare on March 29 2014 (SOL2014-03-29T17:48) by the {\it Interface Region
Imaging Spectrograph} (IRIS). IRIS detected large intensity enhancements of the
\MgII\ {\it h} and {\it k} lines, subordinate triplet lines, and several other metallic
lines at the flare footpoints during this flare. 
We have used the advantage of the slit-scanning mode (rastering) of IRIS and performed,
for the first time, a detailed analysis of spatial and temporal variations of the
spectra. Moreover, we were also able to identify positions of strongest hard X-ray (HXR) 
emissions using the {\it Reuven Ramaty High Energy Solar Spectroscopic Imager}
(RHESSI) observations and to correlate them with the spatial and temporal evolution of
IRIS \MgII\ spectra. 
The light curves of the \MgII\ lines increase and
peak contemporarily with the HXR emissions but decay more gradually. There
are large red asymmetries in the \MgII\ {\it h} and {\it k} lines after the flare peak. 
We see two spatially well separated groups of \MgII\ line profiles, non-reversed and reversed. 
In some cases, the \MgII\ footpoints with reversed profiles are correlated with HXR sources.
We show the spatial and temporal behavior of several other
line parameters (line metrics) and briefly discuss them.  
Finally, we have synthesized the \MgII\ {\it k} line using our non-LTE code with the Multilevel Accelerated Lambda Iteration   (MALI) technique. Two kinds of 
models are considered, the 
flare model F2 of \citeauthor{Machado1980} ({\citeyear{Machado1980}}, \apj{}, \textbf{242}, 336
) and the
models of \citeauthor{RC1983} ({\citeyear{RC1983}}, \apj{}, \textbf{272}, 739,  RC).
Model F2 reproduces the peak intensity of the unreversed \MgII\ {\it k} profile at flare maximum but
does not account for high wing intensities. On the other hand, the RC models 
show the sensitivity of \MgII\ line intensities to various electron-beam parameters. 
Our simulations also show that the microturbulence 
produces a broader line core, while the intense line wings are caused 
by an enhanced line source function.

\end{abstract}

%

\end{opening}
%
\section{Introduction}

The {\it Interface Region Imaging Spectrograph} \citep[IRIS: ][]{DePontieu2014} was launched in the middle of 2013 
and since that time many excellent images and spectra were acquired and analyzed, see \eg\ a special issue of Science in 2014 (Vol. \textbf{346}). 
Among the targets of IRIS, solar flares represent a new challenge because of the
unprecedented spatial resolution achieved in the UV/EUV and potentially high cadence, which allows continuous
rastering of the flaring region. Many flares have been already captured by IRIS 
and some of them are X-class flares. IRIS records spectra in two different wavelength channels: 
FUV (far-UV) from 1332~\AA\ to 1358~\AA\ and 1389~\AA\ to 1407~\AA, and NUV (near-UV) from 2783~\AA\ 
to 2835~\AA. In this article, we concentrate on NUV observations with the objective of studying the
behavior of \MgII\ lines, which appear very bright during flares. To our knowledge, 
Before the launch of IRIS, the only comprehensive study
of \MgII\ lines in a flare was based on the 8th {\it Orbiting Solar Observatory} (OSO-8) spectral observations obtained in hydrogen, calcium and magnesium
lines by the {\it Laboratoire de Physique Stellaire et Planetaire} (LPSP) UV spectrometer \citep[]{Lemaire1984}. Six resonance lines were detected and 
their temporal evolution described: hydrogen L$\alpha$ and L$\beta$, Ca~{\sc ii} H (together with the
blending H$\epsilon$ Balmer line) and K lines, and \MgII\ {\it h} and {\it k} resonance lines. During the observed flare,
the LPSP spectrometer was also able to detect an enhanced emission in three subordinate (intersystem) \MgII\ lines,
which have wavelengths around the \MgII\ {\it k} line and appear as absorption lines in the quiet-Sun (QS) spectrum. 
With the observations from IRIS, the \MgII\ lines during an M class flare was studied in detail by \citet{Kerr2015}. They investigated the spatial and temporal behaviour of \MgII\ lines and found there were red shifts of line centroids, line broadening, and blue asymmetries in flare ribbons.
The formation of \MgII\ lines in the QS has been recently thoroughly discussed by \citet{Leenaarts2013.1}, in preparation
for the analysis of the new IRIS data. It is interesting to note that, according to \citet{Lemaire1984},
both \MgII\ {\it h} and {\it k} integrated line intensities are stronger than the hydrogen L$\alpha$ line during the whole
evolution of the flare. 

Although the hydrogen lines, and in particular the Lyman and Balmer lines, have been extensively modeled for various
flare conditions, including \linebreak time-dependent models, \MgII\ lines are largely unexplored. 
They have been modeled
under the QS conditions \citep{Milkey1974, Bocchialini1996, Uitenbroek1997, Leenaarts2013.1,
Avrett2013}, but concerning the flare atmospheres, the only study is that of \citet{Avrett1986} who
used the well-known semi-empirical flare models of \citet{Machado1980} and synthesized the \MgII\ {\it k} line intensities emergent 
from F1, F2 and F3 model atmospheres. \citet{Lemaire1984} used the models F1 and F2
and computed their own synthetic intensities to compare with \MgII\ OSO-8 observations. 

One of the most spectacular events detected by IRIS was the X1-class flare on 29 March 2014 
\citep{Kleint2015ApJ}.
Several other ground and space instruments observed this flare. Strong \MgII\ emission was already reported by
\citet{Heinzel2014}, who studied the accompanying enhancement of the nearby hydrogen Balmer continuum.
In this article we perform a first detailed analysis of \MgII\ lines in this flare, using the temporal
coverage of the flare from its pre-flare period to the decay phase shortly after the flare maximum.
We defined several `metrics' quantities and show how they evolve in time. 
We also perform non-LTE radiative-transfer modeling using the semi-empirical model F2 \citep{Machado1980} as a 
representative atmospheric snapshot 
of the temperature and density structure and compare our synthetic \MgII\ line intensities
with IRIS observations. We use the theoretical energy balance models
of \citet{RC1983} to study how various electron-beam parameters affect the profiles of the \MgII\ lines.

\section{Observations}\label{sec2}

\begin{figure} 
 \centerline{\includegraphics[width=0.9\textwidth,clip=]{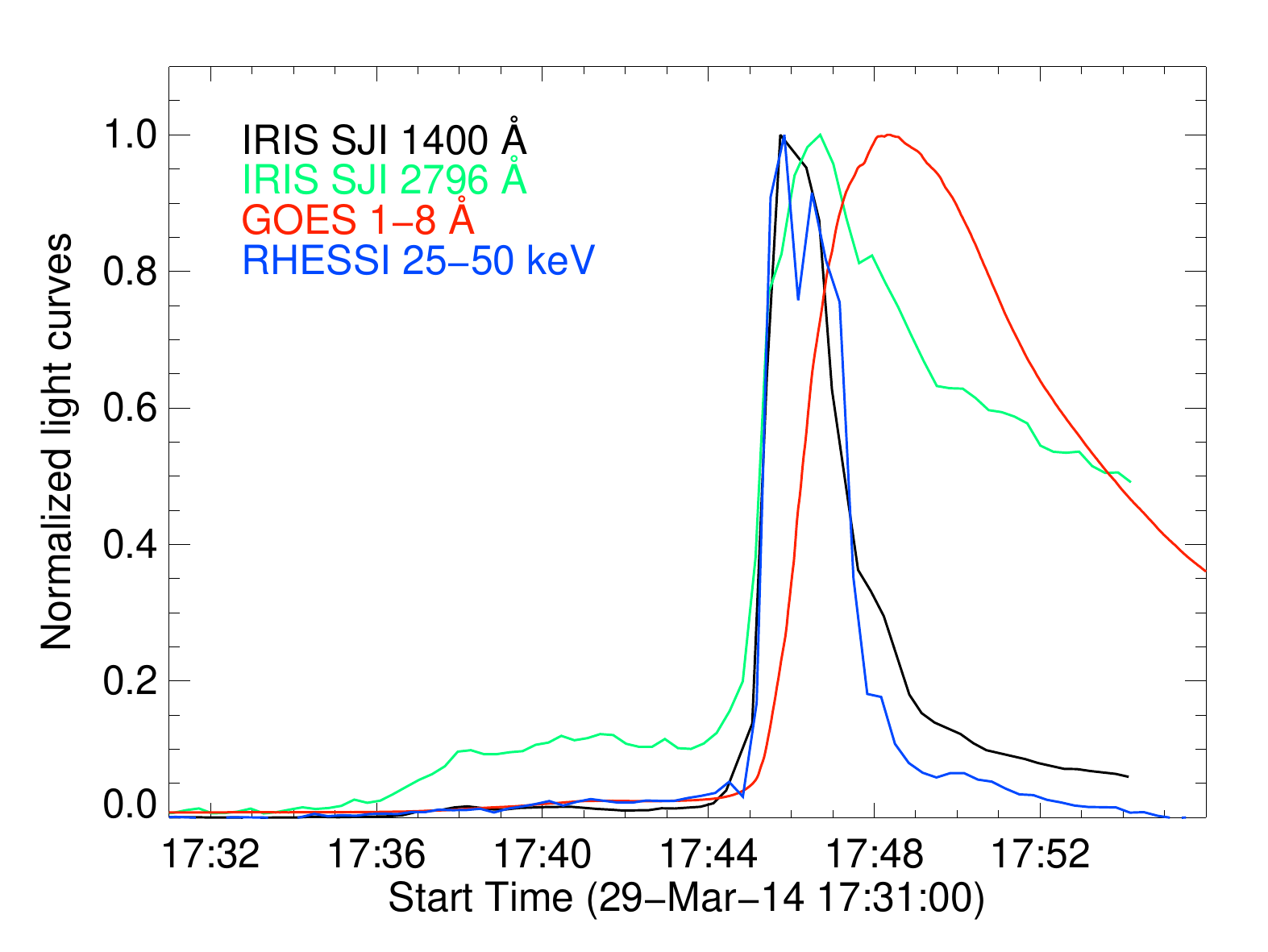}}
 \caption{Normalized and pre-flare subtracted light curves of the 29 March 2014 flare in 1400~\AA\ and 
 2796~\AA\ (the sum of counts in the images of the AR NOAA 12017 taken by the IRIS slitjaw 
 imager 1400~\AA\ and 2796~\AA, divided by the exposure time), in comparison with GOES 
 1 -- 8~\AA\ and RHESSI 25 -- 50~keV HXR light curves.}\label{fig:lightcurve}
\end{figure}

The X1 flare observed on 2014 March 29 occurred in the active region (AR) NOAA 12017. It was observed 
by IRIS from three hours before the flare to a few minutes after the flare peak with a NUV exposure time of 
8.00~s before 17:46:04.78~UT and 2.44~s after 17:46:13.98~UT, and 75~s 
raster cadence  (see Section 2 of \citet{Heinzel2014} and Section 2 of \citet{Young2015} for 
detailed descriptions of IRIS observations of this flare). Light curves of the flare in 
IRIS 1400~\AA\ and IRIS 2796~\AA\ channels, together with X-ray light curves are plotted 
in Figure~\ref{fig:lightcurve}. The light curves in 1400~\AA\ and 
2796~\AA\ are the total counts of the active region, derived from the IRIS slitjaw images, 
divided by the exposure time. Figure~\ref{fig:lightcurve} shows that 
the GOES flux started to rise 
impulsively at 17:45~UT and peaked at 17:48~UT, while fluxes in 1400~\AA, 2796~\AA\ and 
hard X-rays (HXR) peaked 1 -- 3 minutes earlier. Figure~\ref{fig:lightcurve} also 
shows that during the 
decay phase of the flare, the light curves in the 1400~\AA\ slitjaw image followed the 25--50~keV 
HXR fluxes observed by the {\it Reuven Ramaty High Energy Solar Spectroscopic Imager}
\citep[RHESSI: ][]{Lin2002},
while the 2796~\AA\ SJI light curve seems to be slightly delayed and more gradual.

\begin{figure} 
 \centerline{\hspace*{0.015\textwidth}
               \includegraphics[width=0.515\textwidth,clip=]{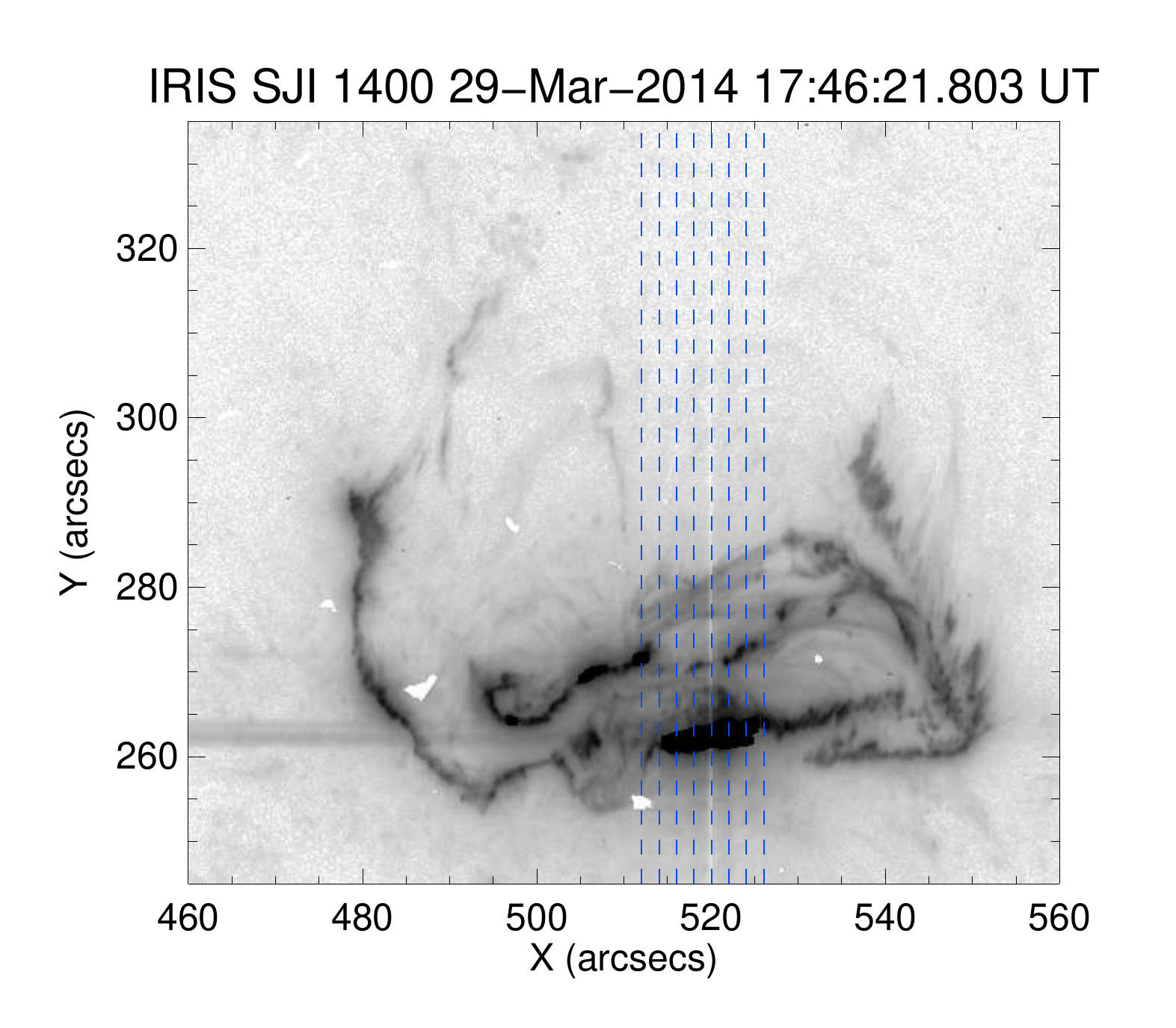}
               \hspace*{-0.03\textwidth}
               \includegraphics[width=0.515\textwidth,clip=]{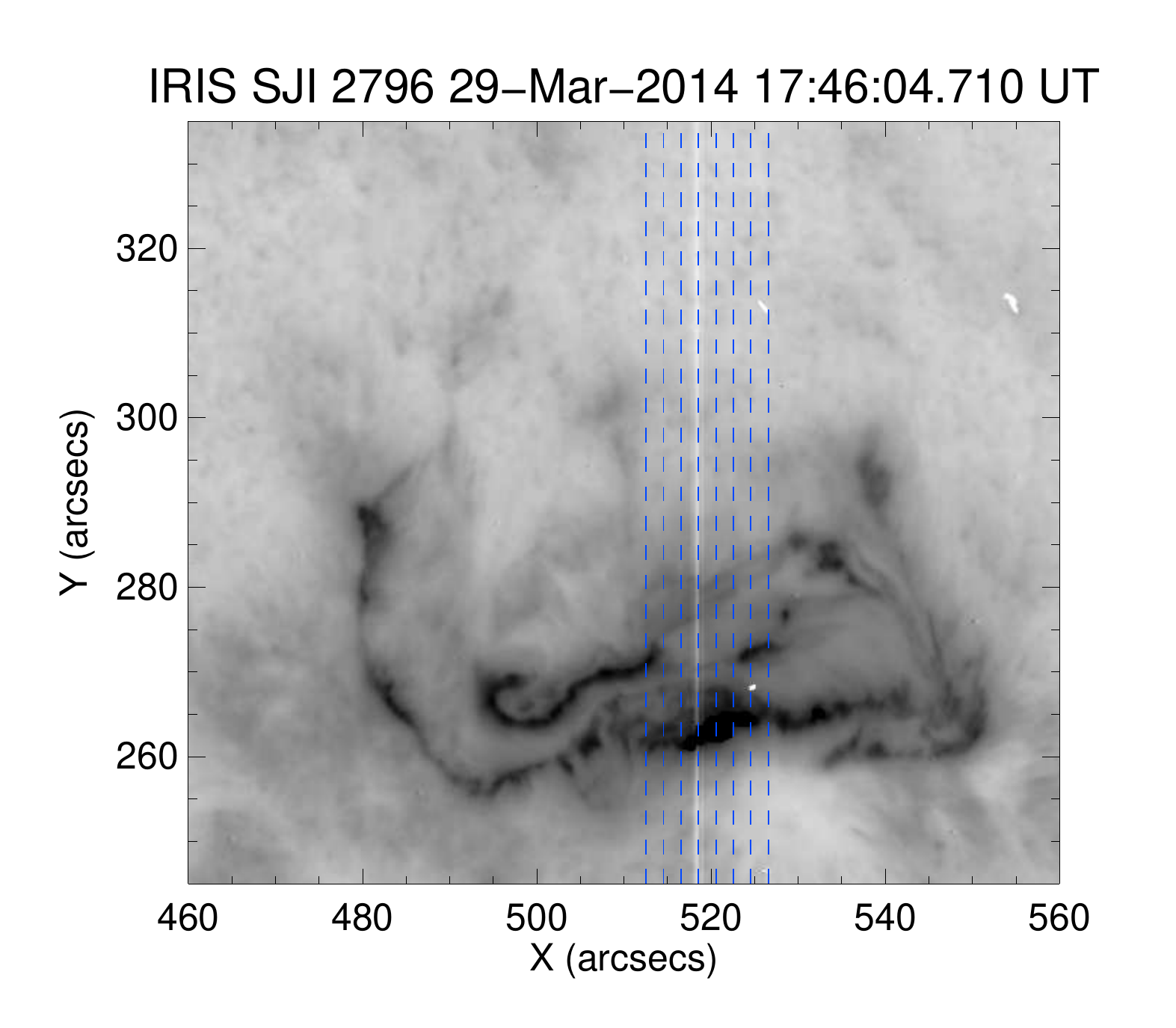}
              }
 \caption{An IRIS 1400~\AA\ slitjaw image taken at 17:46:22~UT (left) and a 2796~\AA\ slitjaw image at 17:46:04~UT (right), with intensity reversed 
 and in log scale. The IRIS slit positions are indicated by vertical dashed lines.}\label{fig:sji}
\end{figure}

Images of the flare at 17:46~UT in 1400~\AA\ and 2796~\AA\ are presented in Figure~\ref{fig:sji}. 
The flare ribbons are clearly seen as black patches in the intensity reversed images. 
Though the contrast of the image in 2796~\AA\ is lower, the flare ribbons  look nearly 
identical in the two images. These images are representative of the plasma in the lower 
transition region (1400~\AA) and upper chromosphere (2796~\AA) \citep{DePontieu2014}.
The IRIS slit positions are also plotted in Figure~\ref{fig:sji} as vertical dashed 
lines; they covered the brightest part of the southern ribbons and part of the 
northern ribbon during the flare. 

\begin{figure} 
 \centerline{\includegraphics[width=0.82\textwidth,clip=]{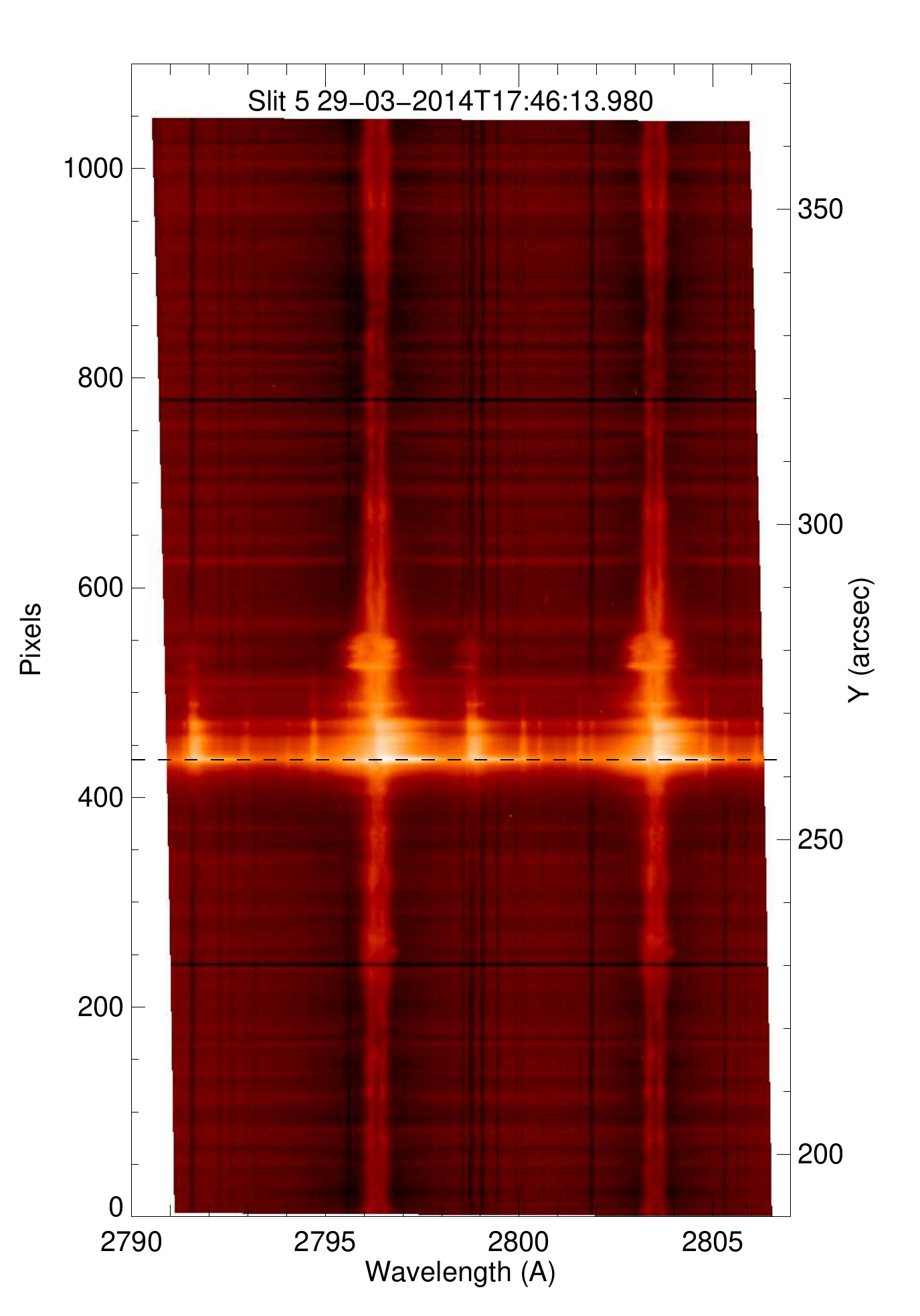}}
 \caption{IRIS spectra of the \MgII\  lines at 17:46:13:98~UT along slit N$^0$ 5. The dashed horizontal line 
 indicates the position of pixel 436, which is used to plot the temporal evolution of the line profiles in Figure~\ref{fig:pixel}.} 
 \label{fig:spec}
\end{figure}

Figure~\ref{fig:spec} shows the IRIS 
spectra of the \MgII\ lines taken at 17:46:13.98~UT along slit N$^0$ 5. 
It is clear that there is strong emission in \MgII\ {\it h}, {\it k}, and the 
subordinate lines, as well as in other metallic lines at 17:46~UT. These emitted 
\MgII\ lines are asymmetric in some of the brightest pixels, with the 
red wings much broader and brighter than the blue wings. The red asymmetries are usually observed 
in chromospheric lines (\eg, H$\alpha$, Ca~{\sc ii} K) during flares 
\citep[\eg,][]{Svestka1962,Tang1983,Ichimoto1984,Asai2012,Deng2013} and are caused by downward motions of the
chromospheric material, which is known as chromospheric condensation \citep{Fisher1985, Fisher1989, Longcope2014}. 
The dashed horizontal line in Figure~\ref{fig:spec} marks the position 
of one footpoint pixel, which is the brightest pixel along the slit N$^0$ 5 at 17:46:13:98~UT and is used to plot 
the temporal evolution of \MgII\ line profiles in Figure~\ref{fig:pixel}.

We use calibrated IRIS level-2 data.  
The intensity observed by IRIS in units of data number (DN) is converted to intensities in 
physical units by the following relation
\begin{equation}
I(\lambda, t)=\frac{n(\lambda, t)~Q(\lambda)~h~c~/~\lambda}{\delta t~\delta\lambda~A_{\rm eff}~\Omega}\ ,
\end{equation}
where $I(\lambda, t)$ is the specific intensity in physical units of 
erg~s$^{-1}$~cm$^{-2}$~sr$^{-1}$~\AA$^{-1}$, $n(\lambda, t)$ 
is the observed intensity in units of DN, read from IRIS level-2 data, 
$\mathbf{Q}(\lambda)$ is the product of the CCD gain and the number of photons needed to create one electron-hole 
pair on the detector, which is 18 photons~DN$^{-1}$ for the NUV band \citep{DePontieu2014}, 
$h$ is the Planck constant in units of erg~s, $c$ is the speed of light in units of cm~s$^{-1}$, 
$\lambda$ is the wavelength in cm, $\delta t$ is the exposure time, which is 8~s for this 
flare in the NUV channel before 17:46~UT, $\delta\lambda$ is 
the spectral resolution, which is 0.02546~\AA\ for the NUV spectra, $A_{\rm eff}$ is the 
effective area for the NUV band obtained from the SolarSoftware \citep[SSW: ][]{Freeland1998} IDL procedure 
{\it iris\_get\_response}, and $\Omega$ is the solid angle subtended by one pixel 
in the y axis of the slit.
For the \MgII\ spectrum between 2790~\AA\ and 2806~\AA, the factor to convert DN~s$^{-1}$ to physical units is in the range of 
$1.82 \times 10^{4}$ to $1.93 \times 10^{4}$. The new IRIS calibration by J.-P. {W{\"u}lser} shows that there is an difference of 10\% for the calibration. 

\begin{figure} 
 \centerline{\includegraphics[width=1.0\textwidth,clip=]{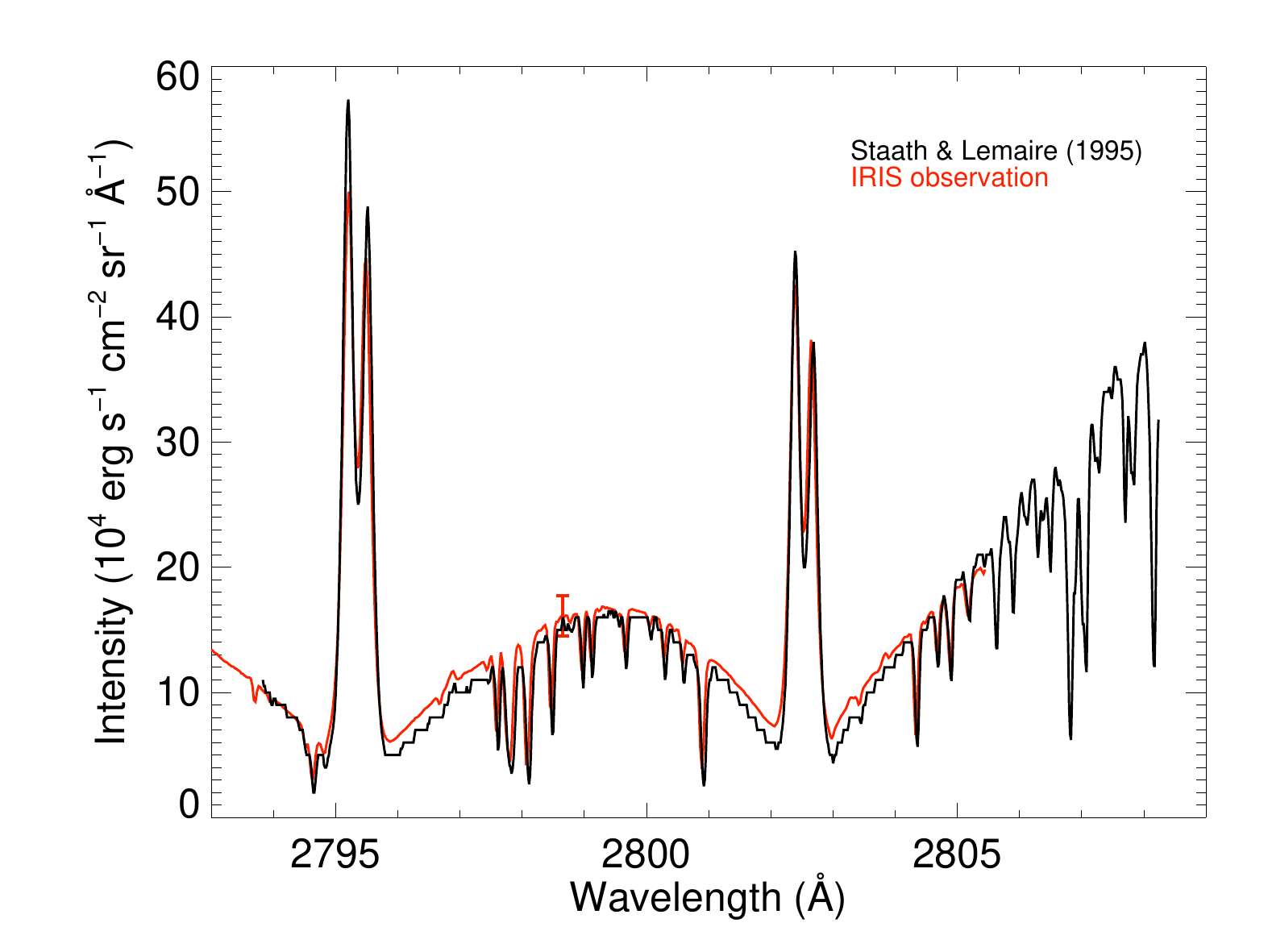}}
 \caption{The \MgII\  spectrum averaged over 100 pixels in the QS region at $\mu=\cos{\theta}$=0.83 in comparison with the QS spectrum observed by RASOLBA at disk center \citep{Staath1995}. Note that IRIS records spectra in vacuum while the RASOLBA observation was made in the air. We shift the IRIS spectrum by 0.84~\AA\ to the blue 
to align the wavelengths of these two observations. The error bar in the figure indicates the 10\% error for the IRIS calibration.}\label{fig:qs}
\end{figure}

To test the absolute radiometric calibration of 
the IRIS NUV spectrum, we plot the QS spectrum in the NUV band observed by IRIS in comparison 
with that observed by the balloon-borne telescope and spectrograph RASOLBA on 19 September 1986 
\citep{Staath1995} in Figure~\ref{fig:qs}. The IRIS QS spectrum shown in Figure~\ref{fig:qs} is a spectrum averaged 
over 100 consecutive pixels (pixels 100--199, corresponding to 16.7\arcsec) and averaged over all eight slit positions 
in the QS region located at $\mu=\cos{\theta}=0.83$ \citep{Heinzel2014} taken around 17:30~UT, while the 
RASOLBA spectrum was an average over 30\arcsec\ slit length and over ten sets of data with alternative 
30~s and 90~s exposure times taken at the center of the solar disk. Figure~\ref{fig:qs} shows that 
the IRIS QS spectrum in the wings of the \MgII\ {\it h} and {\it k} lines is very close to the one from RASOLBA. 
For example, the IRIS QS spectrum at 2800~\AA\ is 
1.62 $\times$ 10$^5$~erg~s$^{-1}$~cm$^{-2}$~sr$^{-1}$~\AA$^{-1}$ while 
the RASOLBA spectrum at the same wavelength is 1.6 $\times$ 10$^5$~erg~s$^{-1}$~cm$^{-2}$~sr$^{-1}$~\AA$^{-1}$. The RASOLBA spectrum at 2800~\AA\ is higher than that from the HRTS-9 rocket \citep{Morrill2008, Pereira2013} and other previous calibrated rocket spectra at QS disk center. For example, \citet{Kohl1976} obtained 14.7$\pm$ 1.8, and  \citet{Bonnet1968} got 15.0$\pm$3.8 \citep[see Table 2a in][for a list of 2800~\AA\ absolute intensity at Sun center]{Lemaire1981}.
In general, the difference between IRIS and RASOLBA QS spectra in the wings is less than 10\% of the RASOLBA data.
The differences in the line cores are more pronounced and they reflect the spatial and temporal variations of the quiet chromosphere.

\begin{figure} 
 \centerline{\includegraphics[width=1.0\textwidth,clip=]{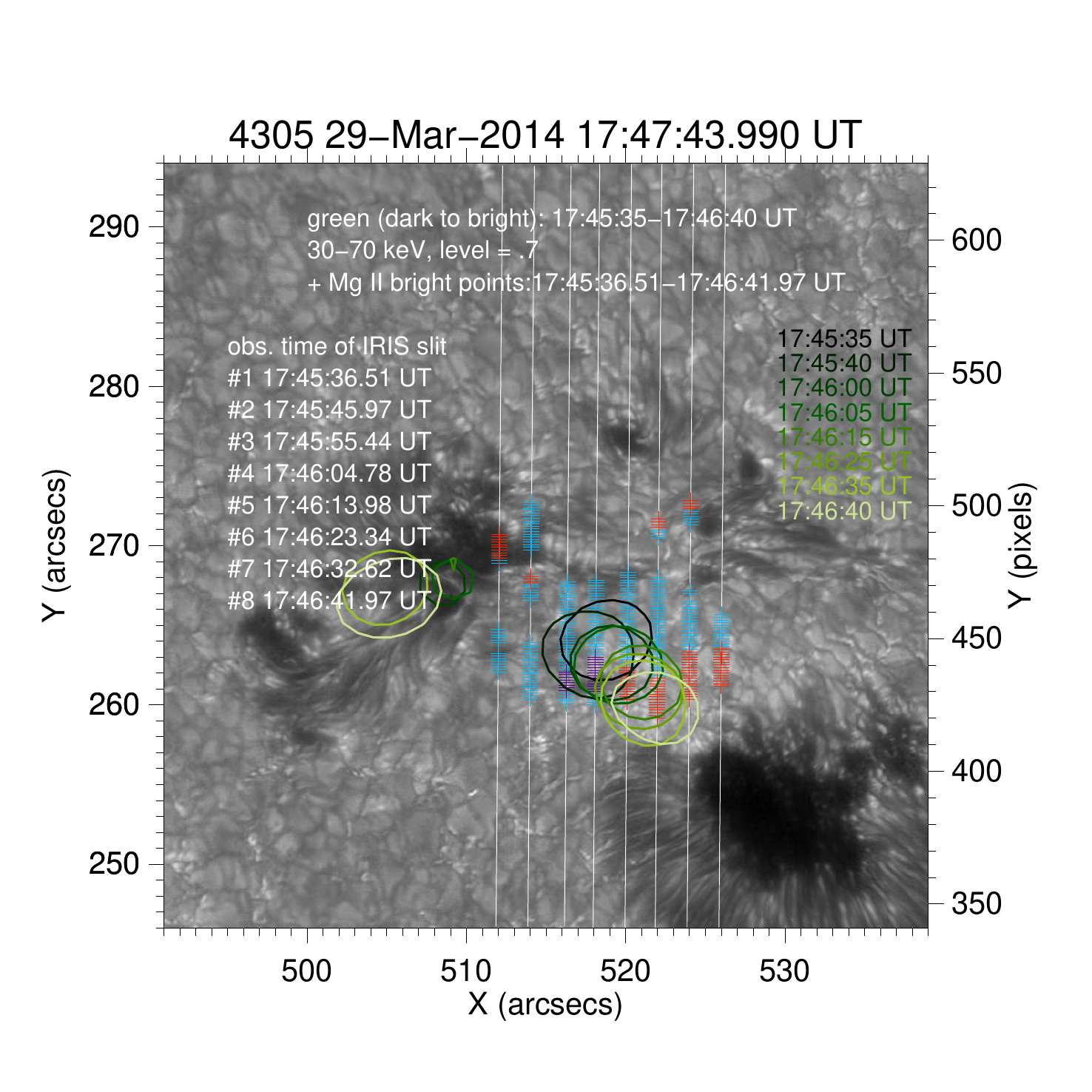}}
 \caption{The \MgII\ bright footpoints (with the integrated intensity for the \MgII\ {\it h}
  line $\geqslant$ 10$^7$~erg~s$^{-1}$~cm$^{-2}$~sr$^{-1}$) along all eight slits and RHESSI 
  30 -- 70~keV HXR contours (level = 0.7) overplotted on a G-band image from the Richard B. Dunn Solar Telecope at the National Solar Observatory (NSO/DST).
  The positions of 
  slits are indicated by vertical white lines. The Mg II bright points are plotted with
 ``+'' sign with blue for a non-reversed line profile, red for a reversed line profile, and violet for
 saturated pixels. The RHESSI contours were created with an integration of 12~s.}\label{fig:rhessi}
\end{figure}
  
To study the relation between the location of footpoints in \MgII\ and HXR sources, 
we plot the \MgII\ bright footpoints along all eight slits for one raster with RHESSI 
30 -- 70~keV HXR contours around the peak of HXR fluxes in Figure~\ref{fig:rhessi}. 
We can see that there are two HXR sources around the peak. 
The IRIS slit did not cross the northern HXR source.
The southern HXR source moved south-west and crossed slits 3--7 from 17:45:35~UT to 17:46:40~UT. 
\MgII\ bright footpoints along slits 5 and 6 with reversed profiles coincide well with 
HXR sources. The \MgII\ line profiles at these footpoints become unreversed after the HXR peak, 
which is clearly seen in the plot of the center-to-peak ratio for slit N$^0$ 5 (Figure~\ref{fig:reversal}). 
Figure~\ref{fig:rhessi} also shows that only a few pixels along slit 3 and 4 (at 17:45:55~UT 
and 17:46:04~UT, respectively) are saturated in the \MgII\ lines. Therefore, the majority of IRIS 
spectra for this flare are of excellent quality.

In Section~\ref{sec3}, we use several parameters (metrics) to characterize the profiles and intensities 
of the \MgII\ lines and to study their variations during the flare. Since the \MgII\ bright footpoints 
along slits 5 and 6 with reversed profiles are correlated with HXR sources at the HXR peak and the 
temporal evolution of these metrics along slit 5 and 6 are very similar, we show the evolution of 
these parameters along slit N$^0$ 5 as an example for most cases.

\section{ Evolution of \MgII\ Spectra during the X1 Flare}\label{sec3}

\begin{figure} 
 \centerline{\includegraphics[width=1.0\textwidth,clip=]{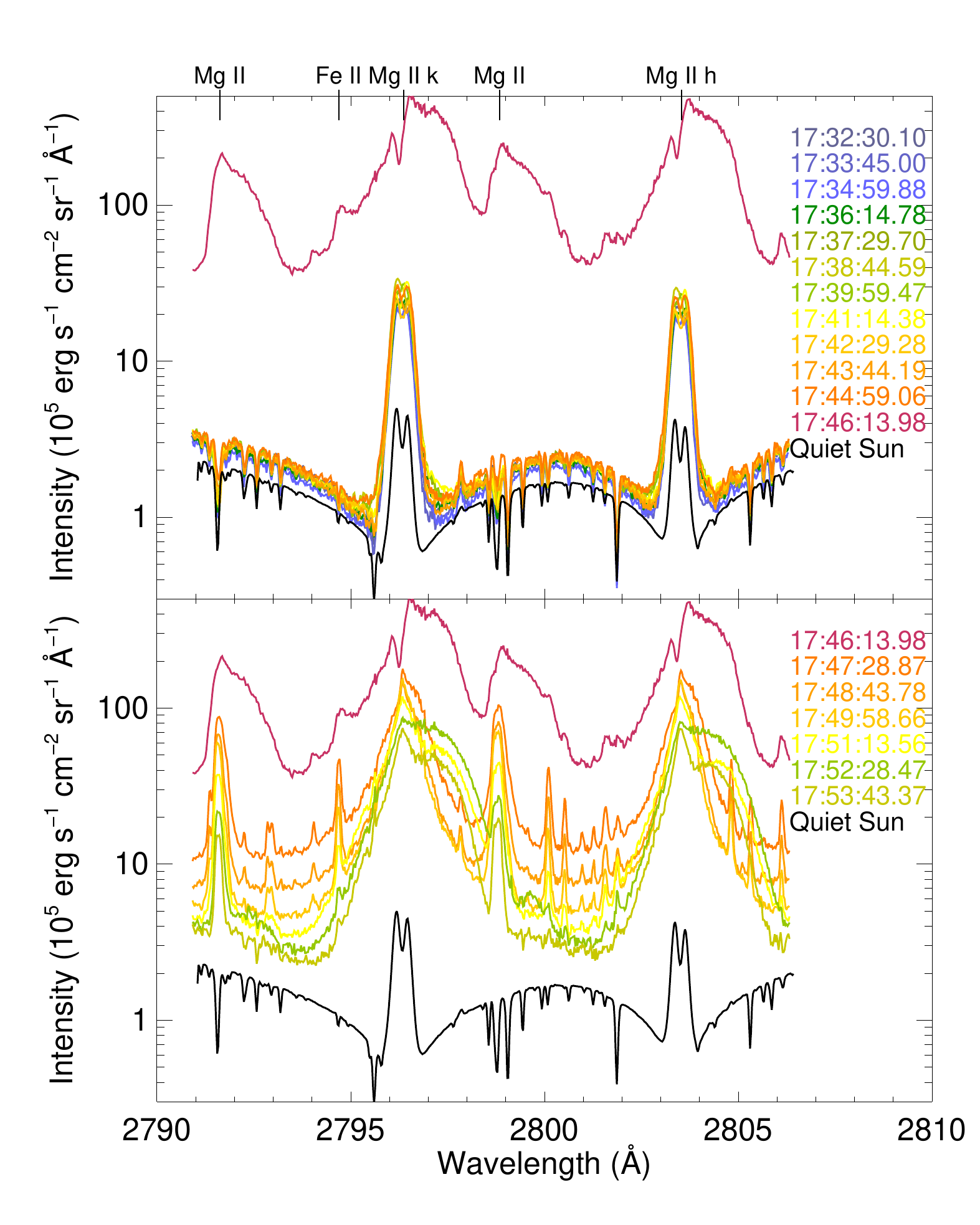}}
 \caption{Temporal evolution of the \MgII\ spectra in one pixel (pixel 436, slit N$^0$ 5) at the footpoint of the 
 flare. The upper panel shows spectra during the rising phase up to the peak time of \MgII\ while the lower panel 
 shows spectra during the decay phase (including the peak time of \MgII). The QS spectrum is plotted 
 in both panels in black as a reference. The position of the pixel is marked by a dashed horizontal 
 line in Figure~\ref{fig:spec} 
 and in Figure~\ref{fig:sub-int} to \ref{fig:fwhm}.}\label{fig:pixel}
\end{figure}

Figure~\ref{fig:pixel} shows the temporal evolution of \MgII\ spectra in one pixel (pixel 436, slit N$^0$ 5) 
at the footpoint of the flare. 
The averaged QS spectrum is also plotted in Figure~\ref{fig:pixel} as a reference. 
Increases of line intensities due to flare energy deposition in the chromosphere are clearly 
seen in Figure~\ref{fig:pixel}. At the center of the \MgII\ {\it h} and {\it k} lines and the subordinate lines (the three 3p -- 3d triplet lines at 2791.6~\AA, 2798.8~\AA, and 2798.8~\AA), 
intensities increase from 17:36~UT, peak at 17:46~UT and decrease thereafter. 
At the peak time (17:46~UT), the intensities in the whole spectral range from 2790~\AA\ to 2806~\AA\ 
increase dramatically, and the widths of the \MgII\ lines and metallic lines also increase. Note that 
for pixel 436 the line profiles of the \MgII\ {\it h} and {\it k} lines are reversed at the peak 
time and become unreversed after the peak. 
For pixels close to pixel 436 the \MgII\ lines are reversed at the peak time, while for positions a few 
pixels north of pixel 436 the \MgII\ lines are in emission and more symmetric at the peak time. After the peak, 
the temporal evolution of the line profiles in \MgII\ bright pixels along slit {N$^0$} 5 is similar to that of pixel 436 
(line profiles are not shown). 
At 17:54~UT, the intensities in the wings of the \MgII\ {\it h} and {\it k} lines return to their pre-flare level, 
which is about 1.5 times the QS intensities. 
The subordinate lines change from absorption to emission at 17:44~UT and remain in emission after the peak time. 
All other metallic lines also go into emission at the peak time (17:46~UT) and come back to absorption at 17:54~UT.
Figure~\ref{fig:pixel} shows that the line profiles of the {\it h} and {\it k} lines are very similar and namely the peak intensities are almost the same.
Significant red asymmetries of the \MgII\ {\it h} and {\it k} lines are observed at the peak time and after 17:51~UT.  
Subordinate lines are also significantly asymmetric at the peak time for pixel 436.

\begin{figure} 
 \centerline{\includegraphics[width=1\textwidth,clip=]{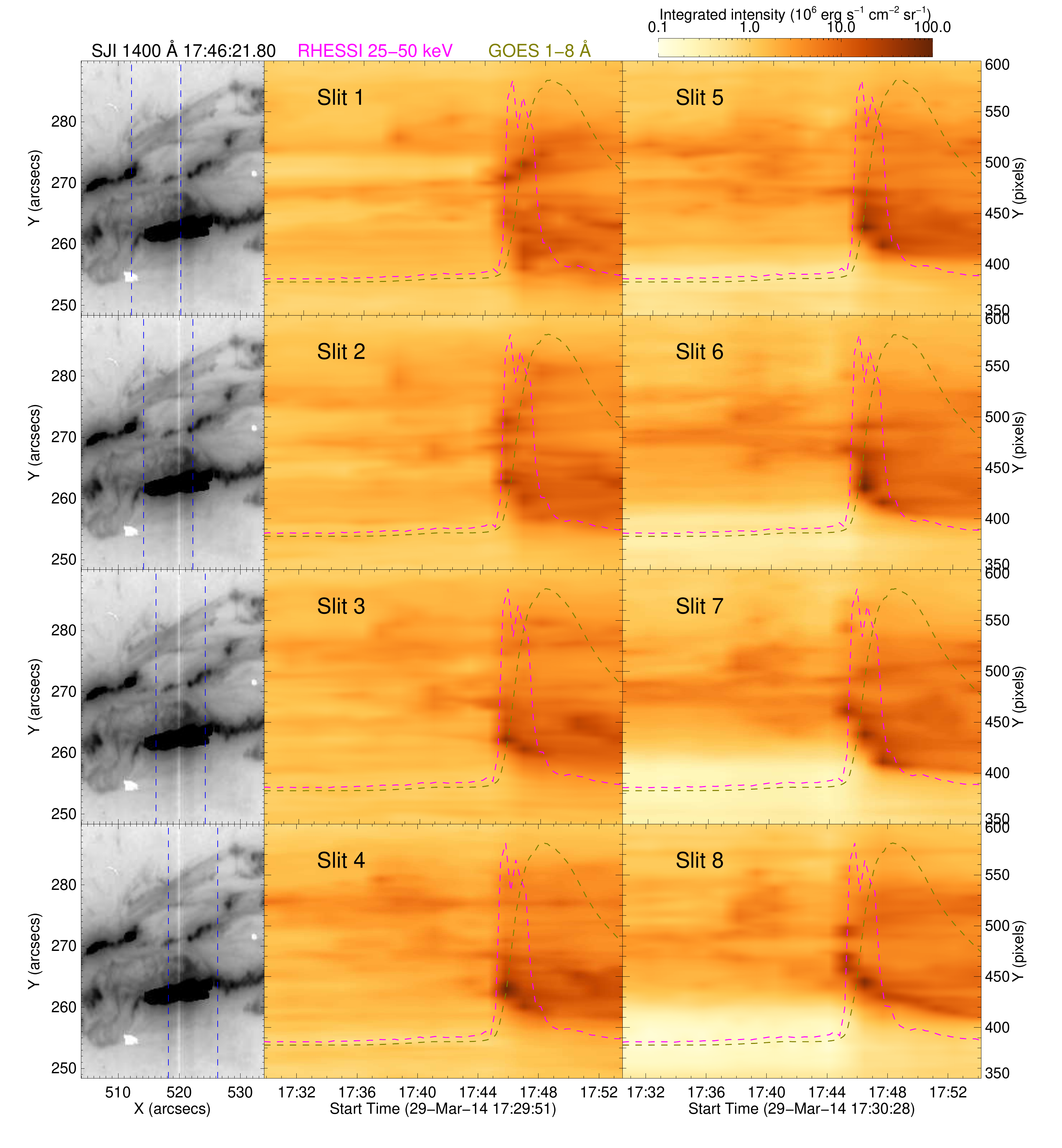}}
 \caption{Left: an IRIS 1400~\AA\ slitjaw image taken at 17:46:21~UT. The vertical dashed lines indicate 
 the positions of the slits which are used to plot the integrated intensity in the center and right columns. 
 Center and right: temporal evolution of the integrated intensities of the \MgII\ {\it h} line over 4~\AA\ along all eight slits. 
  Arbitrarily scaled RHESSI and GOES light curves are also plotted in the center and right panels. 
  Note that there is a tiny shift in the x axes of the integrated intensity, which is due to a 9~s time step between two consecutive  IRIS slit positions.}\label{fig:int}
\end{figure}

The temporal evolution of integrated line intensities for the \MgII\ {\it h} line over 4~\AA\ 
along all eight slits from pixel 350 to pixel 600 is plotted in Figure~\ref{fig:int}.
Also plotted in these figures are the arbitrarily scaled RHESSI 25 -- 50~keV HXR flux in magenta dashed lines and the  GOES 1 -- 8~\AA\ flux in olive dashed lines. The integrated intensities for the \MgII\ {\it h} line in Figure~\ref{fig:int} 
are integrated over a 4~\AA\ band because the {\it h} line from the footpoint pixels is very wide with the full width 
at half maximum (FWHM) around 2~\AA\ after 17:50~UT. Figure~\ref{fig:int} shows that 
the temporal evolution of integrated  \MgII\ {\it h} line intensities at footpoint pixels along all eight slits is 
very similar. Bright pixels with an intense \MgII\ {\it h} line appear when the HXR flux rises impulsively. Then the 
two ribbons move away from each other: the southern ribbon to the south-west and the northern ribbon to the north. The apparent 
motion of flare ribbons in the chromosphere may be caused by the flare energy release moving to new coronal loops.
Figure~\ref{fig:int} shows that there are also some 
variations in the integrated intensities for the \MgII\ {\it h} line at pixels 470--550, which show up before the 
impulsive phase and continue until the end of the IRIS observations. 
The penumbra and umbra in slit 7 and slit 8 respectively are also clearly seen in Figure~\ref{fig:int} with low 
\MgII\ {\it h} line intensities.

\begin{figure}  
 \centerline{\includegraphics[width=1\textwidth,clip=]{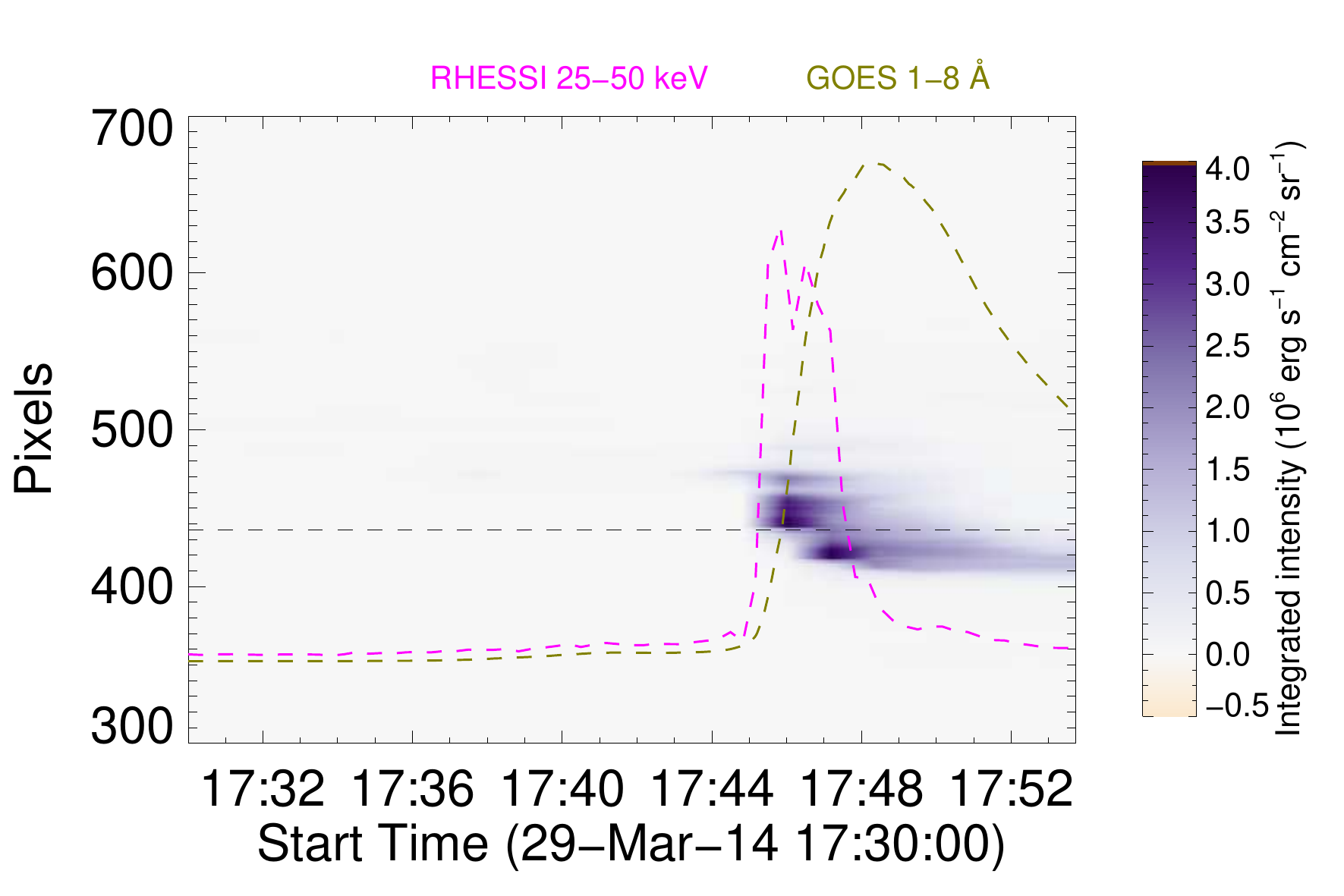}}
 \caption{Temporal evolution of the integrated intensities of the \MgII\ subordinate line at 2791.62~\AA\ over 0.53~\AA\ along slit N$^0$ 5. The dashed horizontal line indicates 
 the position of pixel 436, which is used to plot the temporal evolution of line profiles in Figure~\ref{fig:pixel}.
 Arbitrarily scaled RHESSI and GOES light curves are also plotted in the figure.}\label{fig:sub-int}
\end{figure}

Similar to the \MgII\ {\it h} line, the temporal evolution of subordinate line intensities is very similar at 
footpoint pixels along all eight slits. As an example, we plot the temporal evolution of integrated line intensities for 
the subordinate line at 2791.62~\AA\ over 0.53~\AA\ from pixel 350 to pixel 600 along slit N$^0$ 5 in Figure~\ref{fig:sub-int}.
The \MgII\ subordinate lines are in absorption in QS regions and non-flaring active regions. 
Figure~\ref{fig:sub-int} shows that the subordinate line at 2791.62~\AA\ goes 
into emission at the footpoint pixels during the flare 
with integrated intensities peaking at 17:46--17:47~UT. 
The integrated intensity of the \MgII\ subordinate lines at the footpoint pixels rises 
and peaks as the intensity of the {\it h} line, but decreases faster than the {\it h} line.

\begin{figure}  
 \centerline{\includegraphics[width=1\textwidth,clip=]{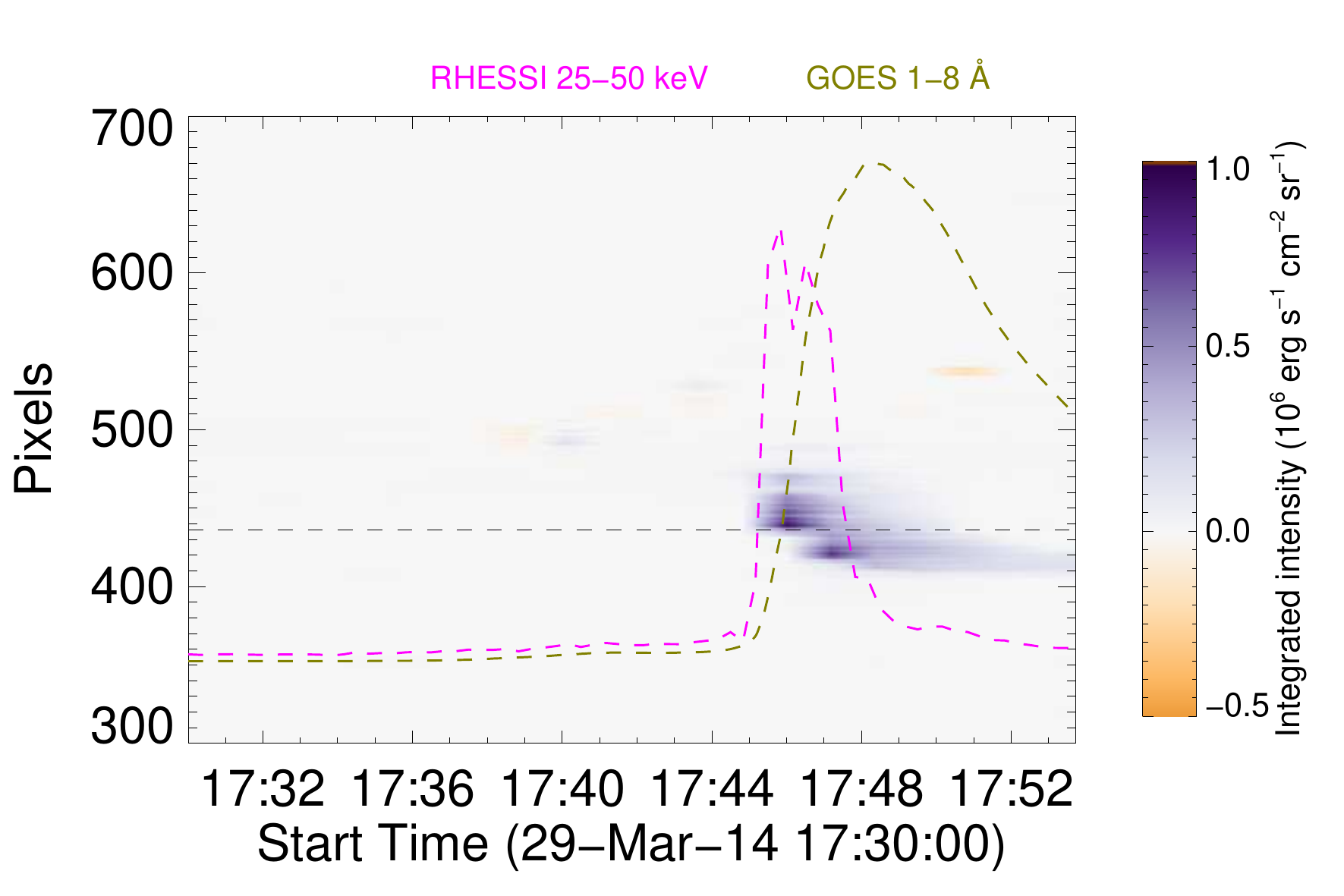}}
 \caption{Temporal evolution of the integrated intensities of Fe~{\sc ii} at 2794.69~\AA\ over 0.64~\AA\ along slit N$^0$ 5. The dashed horizontal line indicates 
 the position of pixel 436, which is used to plot the temporal evolution of the line profiles in Figure~\ref{fig:pixel}. 
 Arbitrarily scaled RHESSI and GOES light curves are also plotted in the figure.}\label{fig:metal-int}
\end{figure}

As seen in Figure~\ref{fig:pixel}, the metallic lines in the wings of the \MgII\ {\it h} and {\it k} lines 
at the footpoint pixels also change from absorption to emission during the flare. The temporal evolution 
of integrated intensities for Fe~{\sc ii} at 2794.69~\AA\ over 0.64~\AA\ along slit N$^0$ 5 is plotted in 
Figure~\ref{fig:metal-int} as an example of those metallic lines. Figure~\ref{fig:metal-int} shows that 
this line goes into emission at 17:45--17:46~UT and returns to absorption at 
17:53:43~UT. Similarly, as for subordinate lines, the integrated intensities for the metallic line also peaks at 
17:46--17:47~UT. The metallic line also decreases faster than the \MgII\ {\it h} line. 
This may be related to the fact that the \MgII\ {\it h} and {\it k} lines are formed higher in 
the chromosphere, while the subordinate lines and the iron line are formed much deeper.
The temporal evolution of Fe~{\sc ii} (2794.69~\AA) intensities at footpoint pixels along other 
slits is similar to that along slit N$^0$ 5 (figures are not shown).

\begin{figure}  
 \centerline{\includegraphics[width=1\textwidth,clip=]{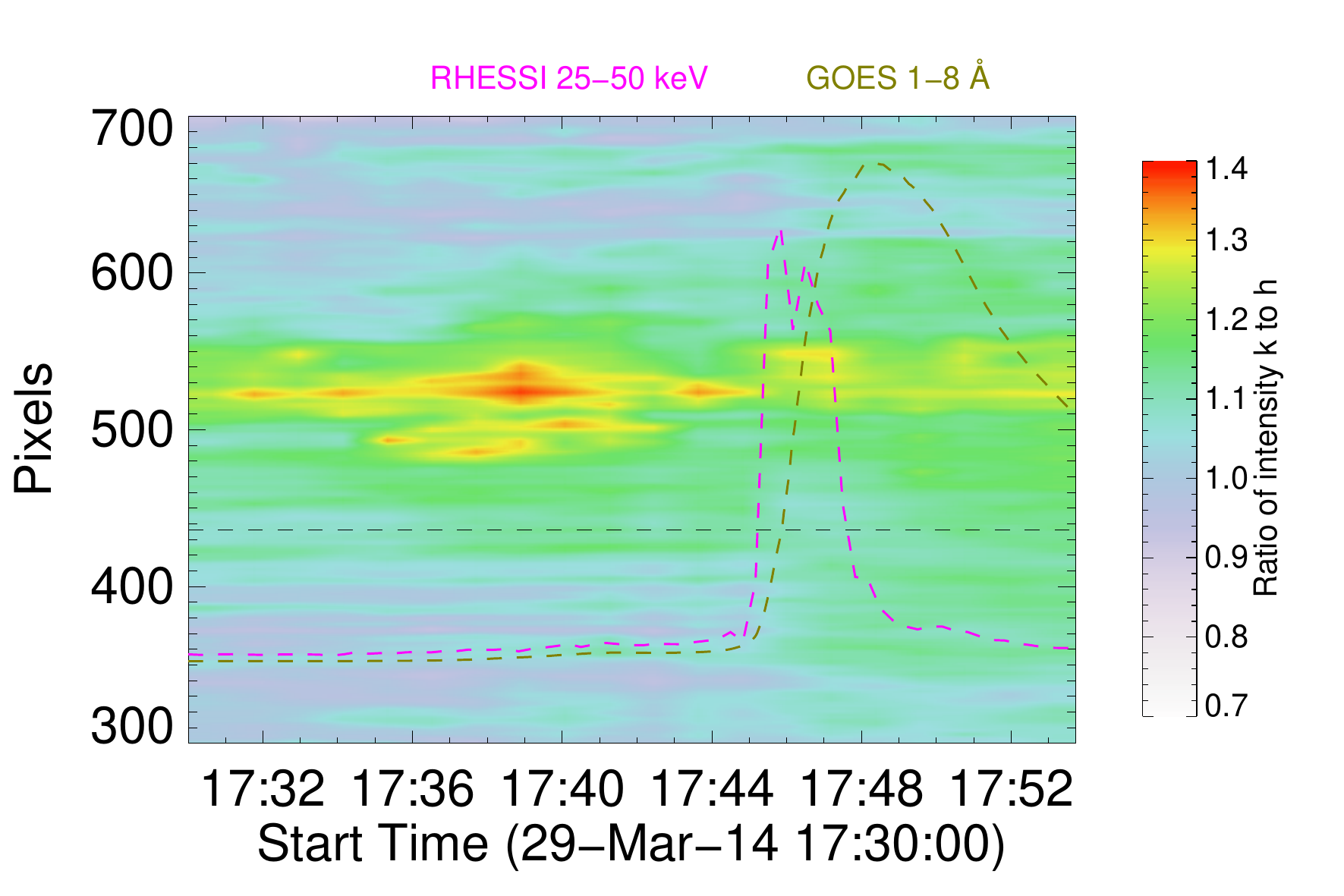}}
 \caption{Ratio of the integrated intensity of the {\it k} line to that of the {\it h} line along slit N$^0$ 5. The dashed horizontal line indicates 
 the position of pixel 436, which is used to plot the temporal evolution of the line profiles in Figure~\ref{fig:pixel}.
 Arbitrarily scaled RHESSI and GOES light curves are also plotted in the figure.}\label{fig:k2h}
\end{figure}

In Figure~\ref{fig:k2h}, we plot the temporal variations of the {\it k} to {\it h} line ratio, which is 
the ratio of integrated intensities 
along slit N$^0$ 5 from pixel 350 to pixel 600 which covers the ribbons and part of QS regions. 
Figure~\ref{fig:k2h} shows that the {\it k} to {\it h} ratio is 
in the range between 0.9 to 1.4 along slit N$^0$ 5. 
For pixel 350 to pixel 600 along all eight slits, the {\it k} to {\it h} ratio is between 0.7 to 1.4.
The ratio at footpoint pixels along all eight slits is around 1.1 (with a standard deviation, 
$\sigma$ = 0.068) and does not change 
during the flare. This is consistent with previous flare observation by OSO-8/LPSP \citep{Lemaire1984} 
and an M class flare observed by IRIS and analyzed by \citet{Kerr2015}, who found the k to h ratio is 
1.07 to 1.19 in flaring pixels and does not change much during flare. 
Figure~\ref{fig:k2h} shows that there are variations of the {\it k} to {\it h} ratio at pixels 470--550 
before the impulsive phase of the flare and during the flare.  

\begin{figure} 
 \centerline{\includegraphics[width=1.0\textwidth,clip=]{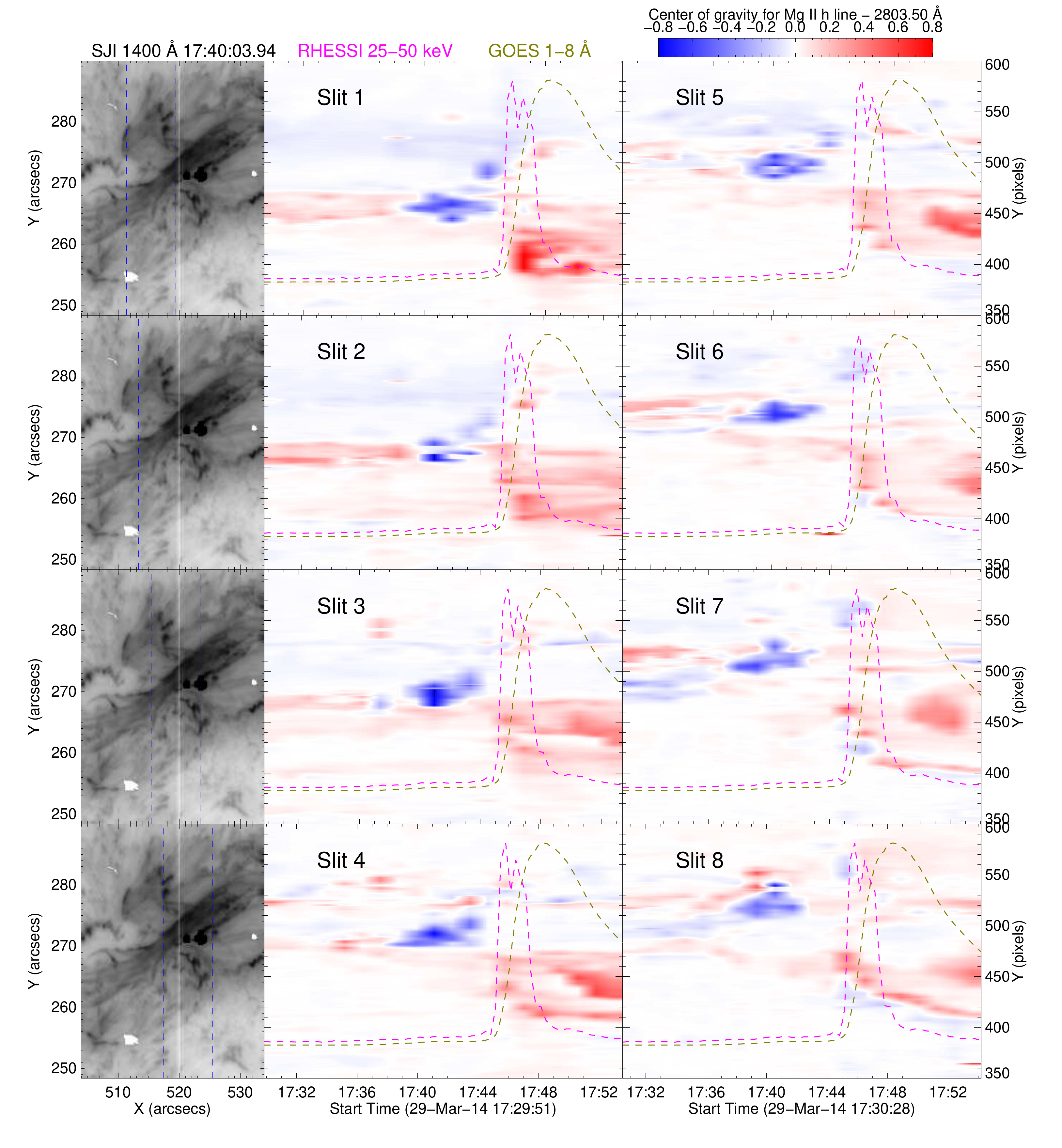}}
 \caption{Left: IRIS 1400~\AA\ slitjaw image taken at 17:40:04~UT. The vertical dashed lines indicate 
 the positions of the slits. For these positions the temporal evolutions of the centers of gravity of the \MgII\ {\it h} line
 are plotted in the center and right columns. 
 Arbitrarily scaled RHESSI and GOES light curves are also plotted in the figure. 
 Note that 2803.50~\AA\ is taken as the reference wavelength for plotting the center of gravity because it is 
 the wavelength with the largest number of pixels in the histogram of the center of gravity for all pixels 
 along the eight slits from 17:29:22~UT to 17:54:11~UT. }\label{fig:lambda-g}
\end{figure}

We use the normalized first-order moment \citep[\eg][]{Druckmuller2007}, which is also known as the center of gravity, to measure the centroid of the line 
\begin{equation}
 \lambda_g(\lambda_0-\Delta\lambda, \lambda_0+\Delta\lambda) = \frac{\int_{\lambda_0-\Delta\lambda}^{\lambda_0+\Delta\lambda} \lambda\; I(\lambda)~{\mathrm d}\lambda}{\int_{\lambda_0-\Delta\lambda}^{\lambda_0+\Delta\lambda} I(\lambda)~{\mathrm d}\lambda} ,
\end{equation}
where $\lambda_0$ is the theoretical line center, $I(\lambda)$ is the intensity at wavelength $\lambda$, $[\lambda_0-\Delta\lambda, \lambda_0+\Delta\lambda]$ is the interval of integration and $\Delta\lambda$ is 
estimated as twice the distance between the peak intensity and 1/$e$ of the peak. 
To reduce the influence of the location of the interval of integration when the observed 
line is shifted substantially from its theoretical center, 
the center of gravity is calculated iteratively \citep{Druckmuller2007}. It is obvious that the center 
of gravity will move to longer wavelengths if there is a red shift and/or red asymmetry of the line and 
it will move to shorter wavelengths due to a blue shift and/or blue asymmetry. Therefore, we plot variations of  
the center of gravity for the \MgII\ {\it h} line along all eight slits in red and blue colors in Figure~\ref{fig:lambda-g}.  
We show that the \MgII\ {\it h} line at footpoint pixels is red shifted and/or red asymmetric when these pixels are bright. 
The center of gravity for pixels at the northern ribbon is less shifted compared to pixels in the 
southern ribbon and is only shifted at the time when the pixel is brightest.
For most pixels in the southern ribbon, the center of gravity is also shifted to the red wing after 17:49~UT (peak of GOES light curve).  
The profiles of the \MgII\ {\it h} line at these footpoint pixels 
show that the shift is mainly caused by a red asymmetry and therefore the line width is larger at the same time. 
Figure~\ref{fig:lambda-g} also shows that there are large shifts to the blue at some pixels before the impulsive 
phase of the flare. Slitjaw images show that the blue shifts and/or blue asymmetries at these pixels are caused by an erupting filament \citep{Kleint2015ApJ}.
There are also some small shifts at pixels not related to the filament and flare. Compared to the map of the 
$k$ to {\it h} ratio for slit N$^0$ 5 in Figure~\ref{fig:k2h}, these shifts may be related to a larger {\it k} to {\it h} ratio. 

\begin{figure} 
 \centerline{\includegraphics[width=1\textwidth,clip=]{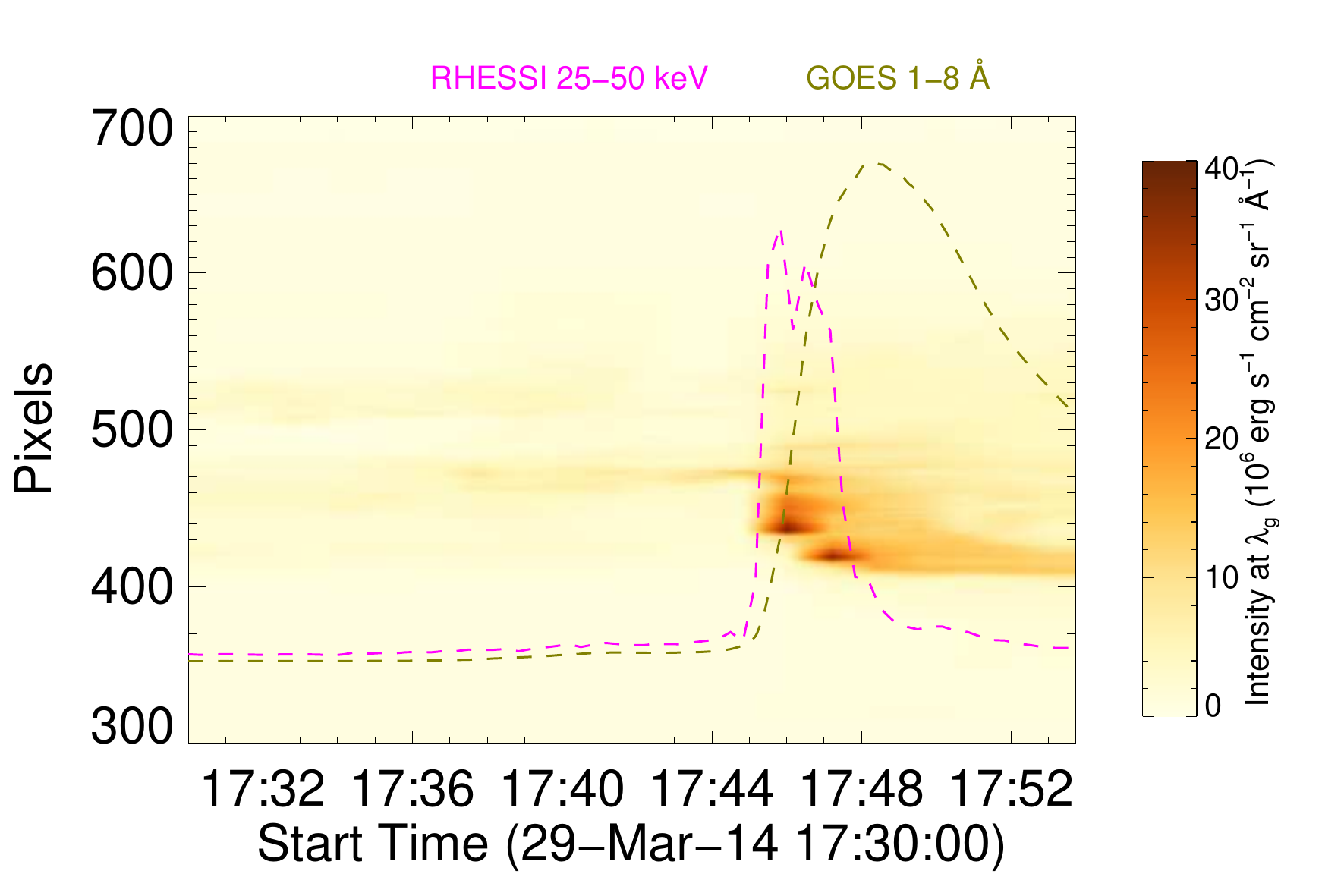}}
 \caption{\MgII\ {\it h} line intensity at its center of gravity along slit {N$^0$} 5. The dashed horizontal line indicates 
 the position of pixel 436, which is used to plot the temporal evolution of line profiles in Figure~\ref{fig:pixel}.
 Arbitrarily scaled RHESSI and GOES light curves are also plotted in the figure.}\label{fig:I-c}
\end{figure}

In Figure~\ref{fig:I-c}, we plot the variations of the \MgII\ {\it h} intensity at its center of gravity along slit N$^0$ 5. 
The variations are mainly at footpoint pixels with intensities increasing at 17:45--17:46~UT and peaking at 17:46--17:47~UT. 
For other slits, the map of the \MgII\ {\it h} intensity at its center of gravity also shows clearly when the footpoint
pixels start to brighten and become brightest (figures are not shown). 

\begin{figure} 
 \centerline{\includegraphics[width=1\textwidth,clip=]{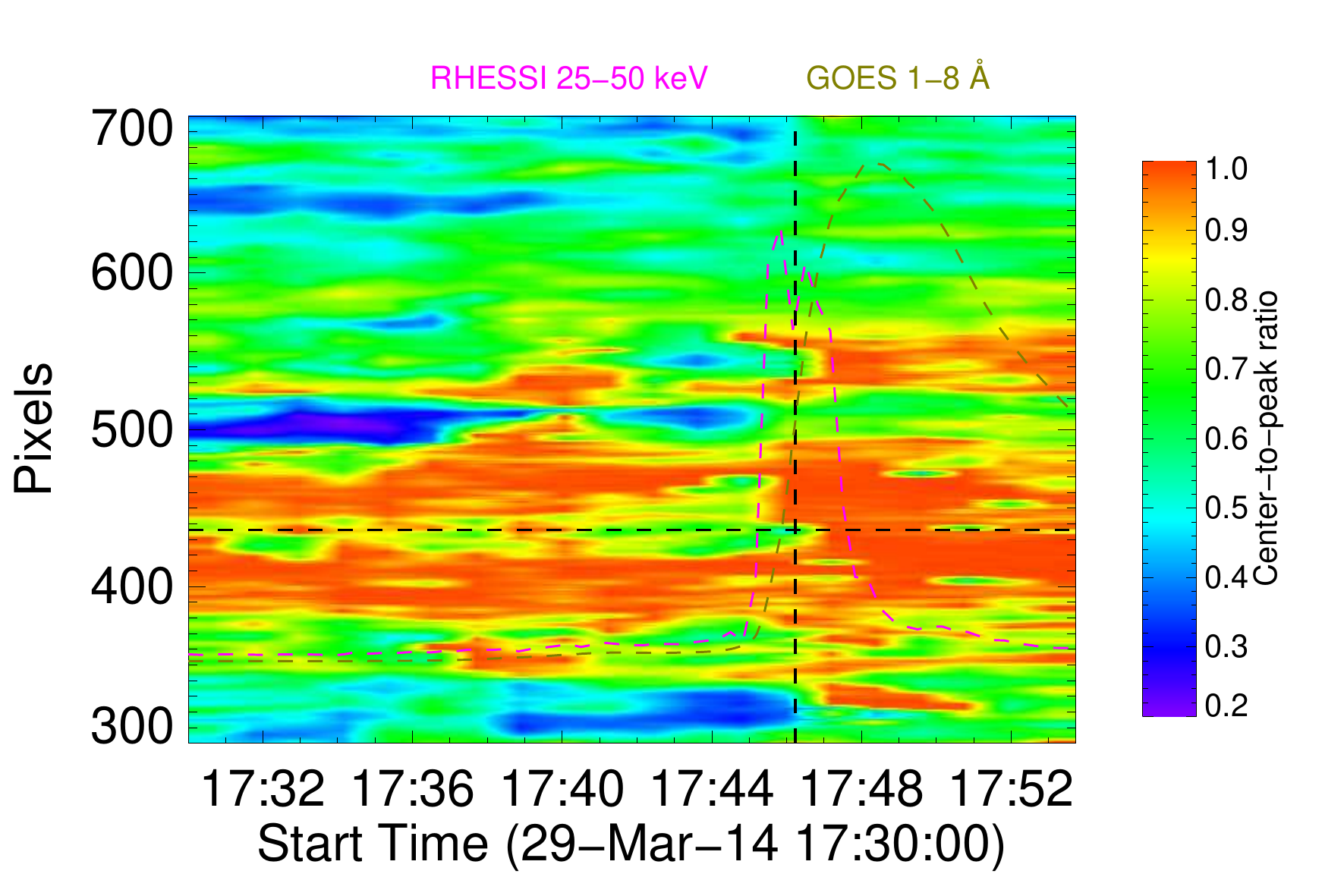}}
 \caption{Center-to-peak ratio of the Mg {\it h} line along slit N$^0$ 5. The dashed horizontal line indicates 
 the position of pixel 436, which is used to plot the temporal evolution of line profiles in Figure~\ref{fig:pixel}. 
 The vertical dashed line indicates the time 17:46:13:98~UT when the \MgII\ bright footpoints along 
 slit N$^0$ 5 are plotted in Figure~\ref{fig:rhessi}.
 Arbitrarily scaled RHESSI and GOES light curves are also plotted in the figure.}\label{fig:reversal}
 \end{figure}

To describe how much the cores of the \MgII\ lines are reversed, we define the center-to-peak ratio as 
$I_{\rm c}/I_{\rm p}$, where $I_{\rm c}$ is the intensity at the reversal minimum and $I_{\rm p}$ is 
the intensity at the main peak of the line, which is the maximum intensity of the line. 
Therefore, the ratio is smaller for line profiles with deeper reversal and larger for line profiles
which are less reversed. The ratio value is equal to one for a pure emission line. The variations of this ratio 
for the \MgII\ {\it h} line along slit N$^0$ 5 are plotted in Figure~\ref{fig:reversal}. 
We can see that the ratio at some footpoint pixels is close to one, while at other pixels it is about 0.7 
(for example, pixel 436, whose profile is shown in Figure~\ref{fig:pixel}). 
For the filament, the center-to-peak ratio is very small before its eruption and increases when erupting.
However, it is interesting to observe a ratio close to one in some active region pixels 
also before the flare (for example, pixels 380 to 420 and pixels 450 to 470 along slit N$^0$ 5).
 
\begin{figure} 
 \centerline{\includegraphics[width=1\textwidth,clip=]{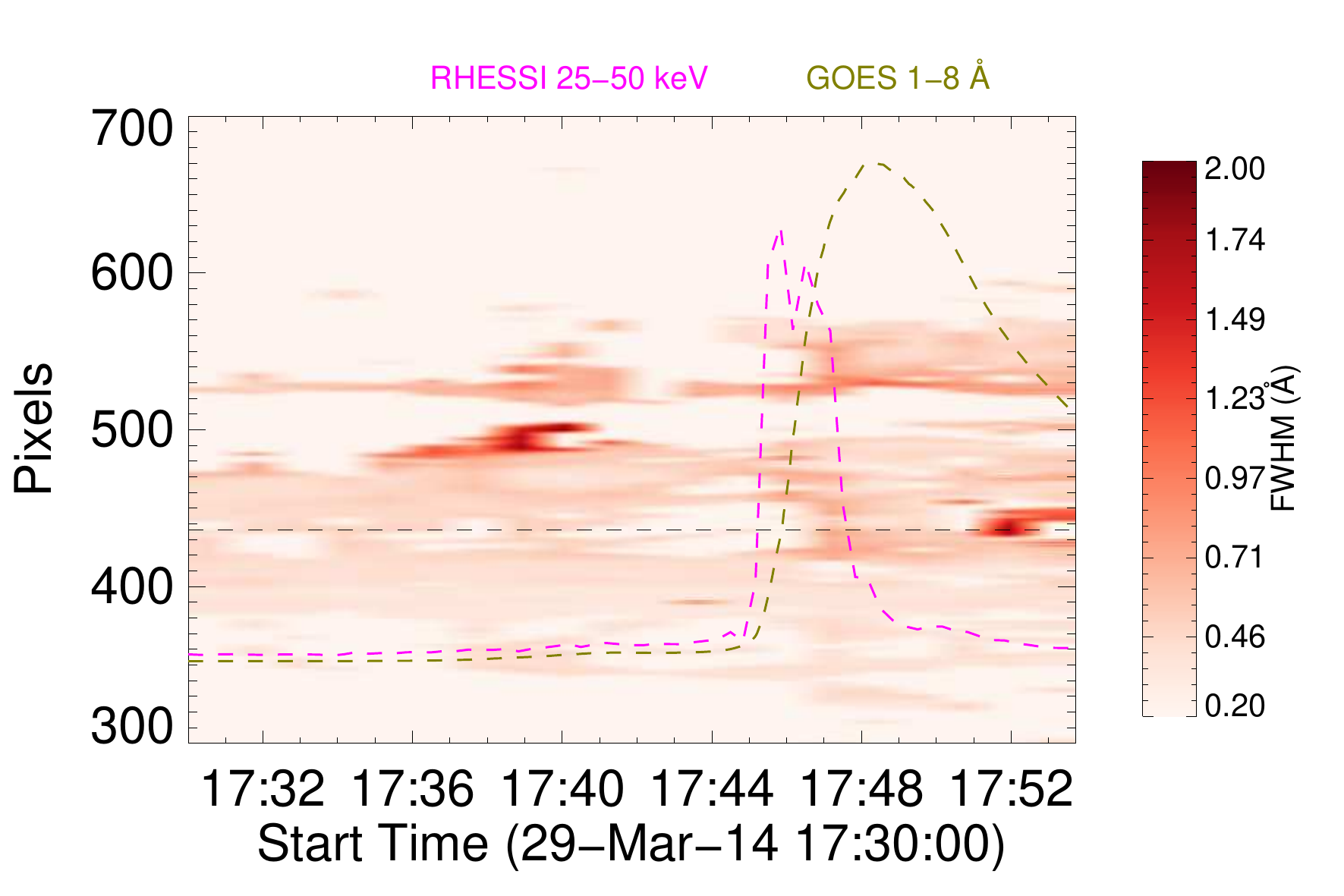}}
 \caption{FWHM of the Mg {\it h} line along slit N$^0$ 5. The dashed horizontal line indicates 
 the position of pixel 436, which is used to plot the temporal evolution of line profiles in Figure~\ref{fig:pixel}. 
 Arbitrarily scaled RHESSI and GOES light curves are also plotted in the figure.}\label{fig:fwhm}
 \end{figure}

To show variations of the line widths during the flare, we plot in Figure~\ref{fig:fwhm} the FWHM of 
the \MgII\ {\it h} line along slit N$^0$ 5 when it appears as an 
emission line (\ie\ at the time when its center-to-peak ratio is larger than 0.8). We see that 
the FWHM of the \MgII\ {\it h} line at the footpoint pixels is larger 
after 17:50~UT when there are red asymmetries in the line. 
For this flare, the maximum FWHM of the \MgII\ {\it h} line is 2~\AA\ which is four times the FWHM at 
disk center QS regions, where it is observed to be 0.5 $\pm$ 0.02\AA\ \citep{Kohl1976, Staath1995}.
Figure~\ref{fig:fwhm} shows that the FWHM of the \MgII\ {\it h} line in the filament pixels increases two or 
three minutes before and during the eruption.
The shape of a line is also characterized by its various order moments. 
The line width is expressed in terms of the second order central moment, which is the square of the standard deviation.  
For an emission line with a Gaussian profile, its FWHM is proportional to its standard deviation.

Comparing with the flare reported by \citet{Lemaire1984}, there are several similarities.
First, there are great great emission enhancements in the \MgII\ {\it h} and {\it k} lines and the subordinate lines. 
Second, the light curves of the \MgII\ {\it h} or {\it k} lines rise and peak at the same time as the subordinate lines.
Finally, the $k$ to $h$ ratio is 1.1 $\pm$ 0.068 at footpoint pixels and does not vary with time during the flare.
Apart from these similarities, we observe red asymmetries at the flare peak and the decay phase for the X-class flare. 
Different from the flare of \citet{Lemaire1984}, in our flare the light curves of the subordinate lines  decrease 
faster than the light curves of the {\it h} and {\it k} lines after the peak time. It is interesting to note that 
some \MgII\ line profiles at some footpoint pixels are reversed, while they are non-reversed at other footpoints  during the flare,
as we have seen before.

\section{Synthetic Spectra}\label{sec4}

In this section, we present preliminary modeling of the \MgII\ lines in a flaring
atmosphere to see whether existing flare models can
quantitatively reproduce the calibrated IRIS spectra. As we have mentioned in the
introduction, the only existing fitting of the \MgII\ lines in a flare was
performed in \citet{Lemaire1984}. 
We solved the non-LTE radiative-transfer 
problem to model the \MgII\  lines 
using a static semi-empirical flare atmosphere.
Our numerical code is based on a 1D plane-parallel geometry and assumes the 
atmosphere in a hydrostatic equilibrium. The non-LTE problem
is treated by the so-called Multilevel Accelerated Lambda Iteration (MALI) technique \citep{Rybicki1991, Heinzel1995, Kasparova2002}. 
By taking the temperature and pressure 
structure given
by the flare atmosphere, we obtained hydrogen-level populations and electron 
densities from a five-level plus continuum atomic model of hydrogen.
Hydrogen Ly $\alpha$ and Ly $\beta$ lines were treated with the partial 
frequency redistribution \citep[see][]{Hubeny2014}. 
Using the fixed electron density structure, 
the radiative transfer problem was solved for a five-level plus continuum atomic 
model of \MgII\ with the two resonance lines {\it k} and {\it h} and the 3p -- 3d subordinate triplet around the
{\it k} line, for details of the atomic model see \citet{Heinzel2014AA}. 

\begin{figure} 
 \centerline{\includegraphics[width=1\textwidth,clip=]{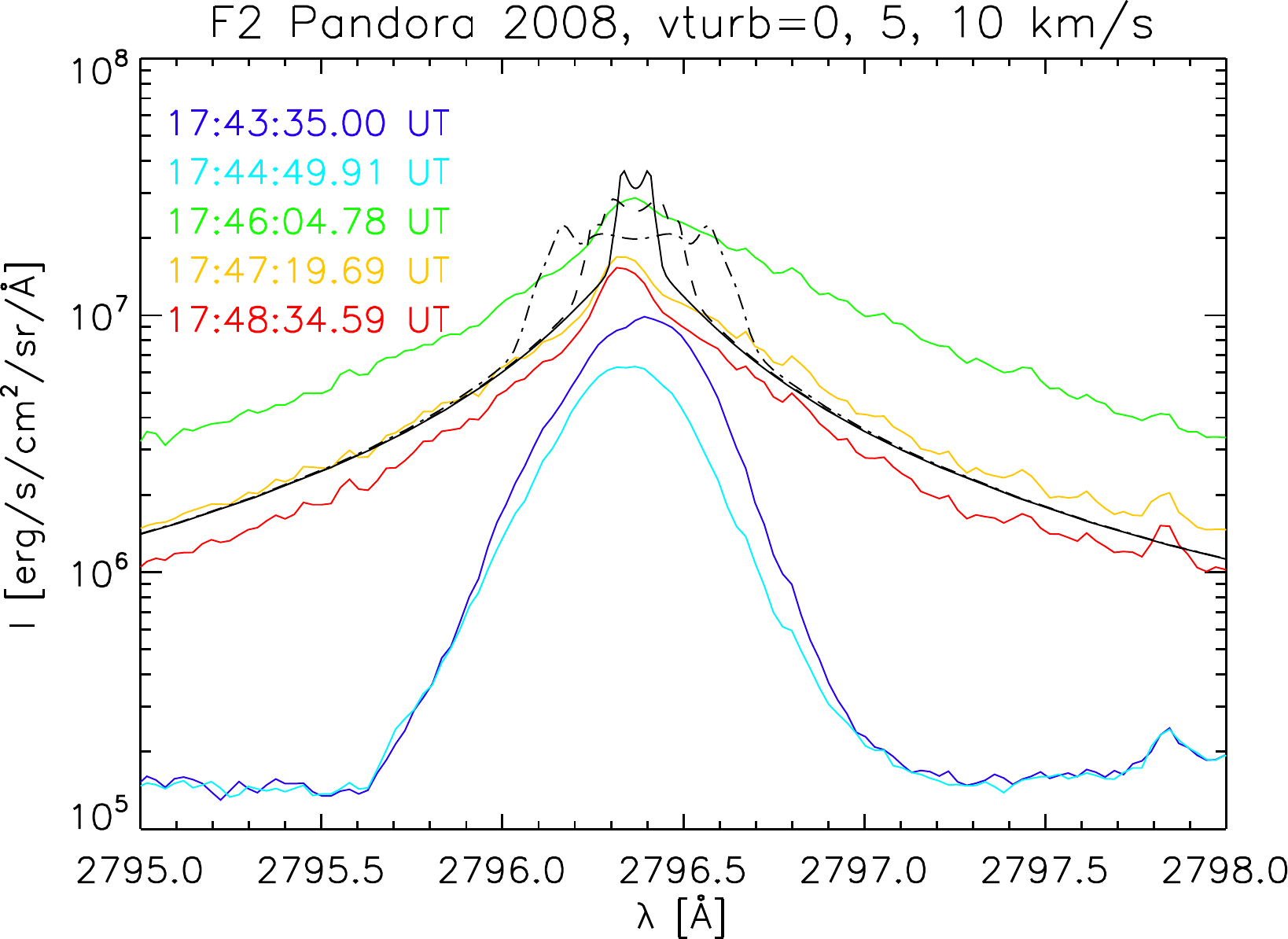}}
 \caption{A comparison between the synthetic \MgII\ {\it k} line intensities computed from model F2
and IRIS observations for slit  N$^0$ 4 in pixel 447. The profiles have been obtained for three microturbulent
velocities: 0 km s$^{-1}$ - full line, 5 km s$^{-1}$ - dashed line, and 10 km
s$^{-1}$ dash-dotted line. `Pandora 2008' is the version of the F2 model 
with a slightly different coronal pressure from the original model of \citet{Machado1980}. 
Note the intensity decrease at time 17:44:49.91.
 }\label{fig:F2}
\end{figure}

\begin{figure} 
   \vspace{-0.01\textwidth}   
   \centerline{\hspace*{0.015\textwidth}
               \includegraphics[width=0.5\textwidth,clip=]{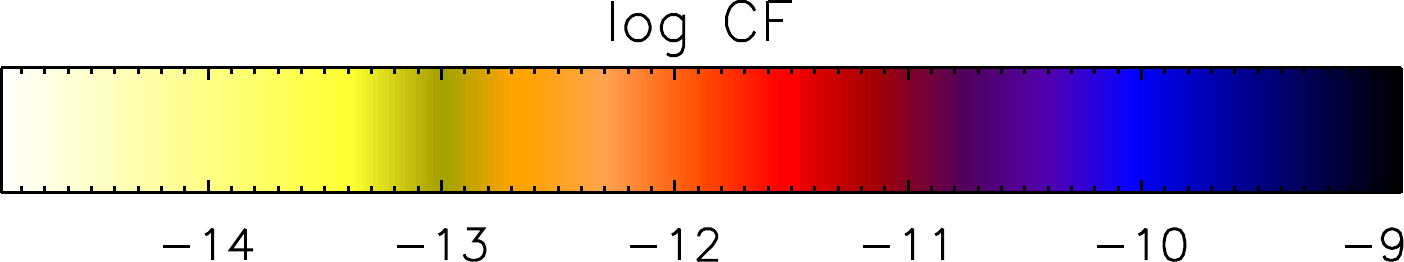}
              }  
   \vspace{0.02\textwidth}   
   \centerline{\hspace*{0.015\textwidth}
               \includegraphics[width=0.5\textwidth,clip=]{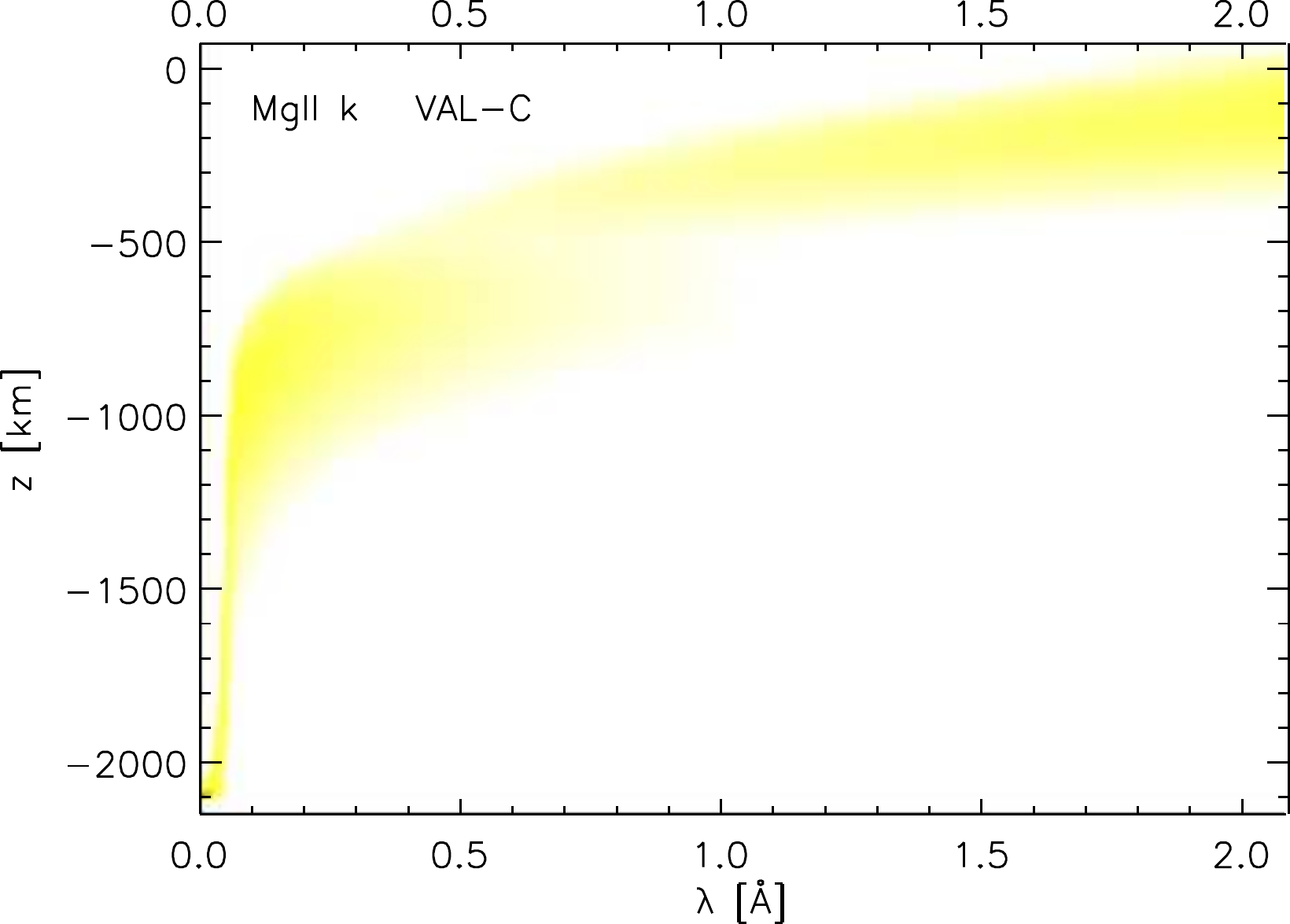}
               \includegraphics[width=0.5\textwidth,clip=]{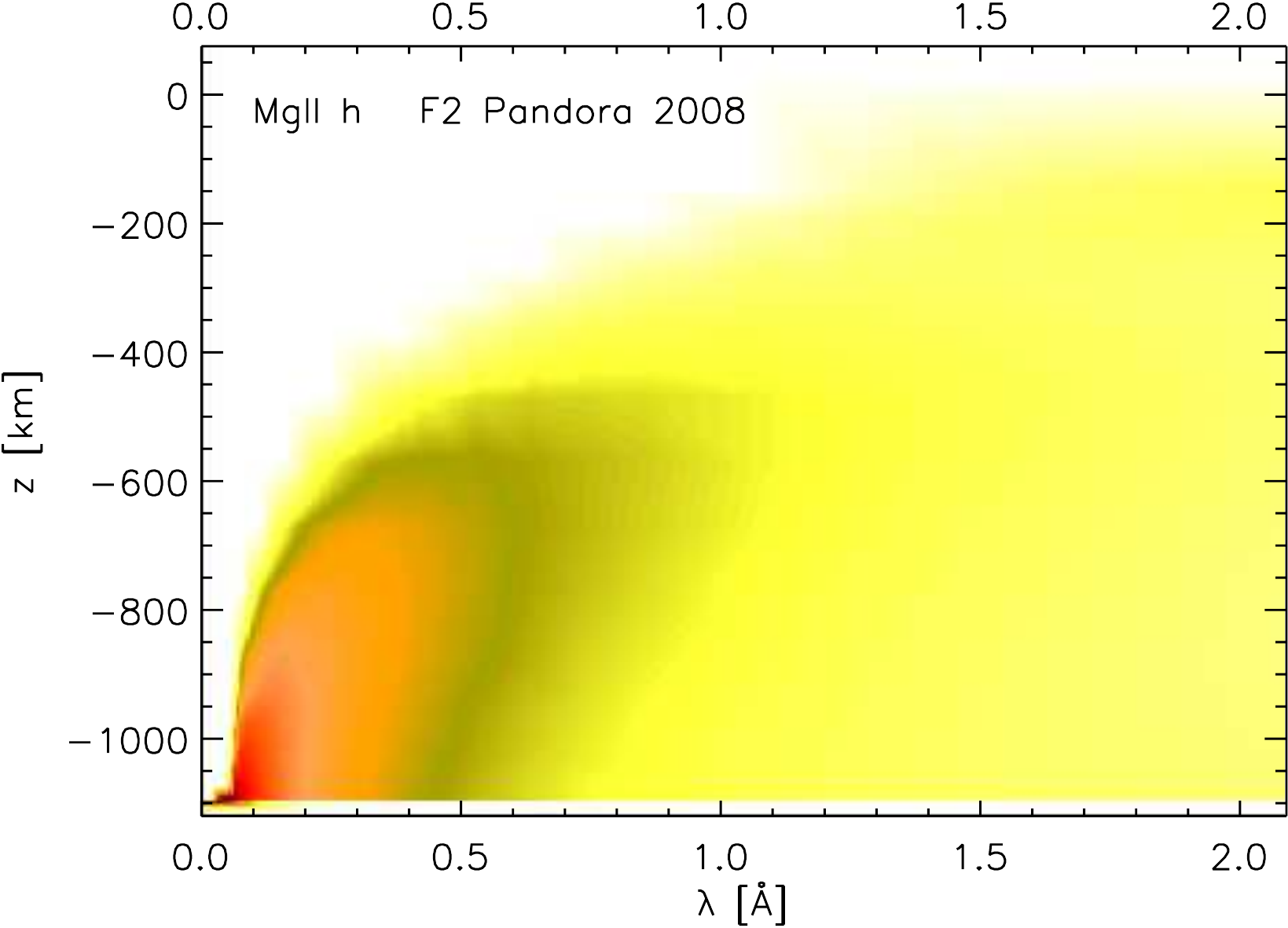}
              }
     \vspace{-0.34\textwidth}   
     \centerline{\Large \bf     
      \hspace{0.4\textwidth}  \color{black}{\small (a)}
      \hspace{0.42\textwidth}  \color{black}{\small (b)}
         \hfill}
     \vspace{0.31\textwidth}    
\caption{\MgII\ {\it k} line contribution function for: (a)  VAL C and (b) F2 model  atmospheres.
Note the height scale $z$ is inverted ($z=0$ denotes photosphere) and it is different for each model. 
We clearly see how the chromosphere is
affected by the heating, contributing significantly to the line wings.
The QS wings are formed much deeper.}
   \label{fig:cf}
\end{figure}

After modeling, we compare our synthetic spectra to the observed IRIS spectra. 
As the flare was of class X1, we chose
the classical semi-empirical flare model F2 of \citet{Machado1980}. 
From an extended grid
of synthetic line profiles as shown in \citet{Avrett1986} 
we noticed that 
the \MgII\ {\it k} line profile corresponding to model F2 is non-reversed, a feature
we observe at several positions.
Its peak intensity is quite comparable to that detected around the
maximum of \MgII\ emission at pixel 447 slit N$^0$ 4, just above the saturation region. 
Using the above mentioned non-LTE 
technique, we computed the \MgII\ line profiles from the F2 model.
The resulting \MgII\ {\it k}
line is shown in Figure~\ref{fig:F2}, where we present a few examples for various values 
of the microturbulent velocity ranging from 0 to 10 km s$^{-1}$. Note that this
microturbulence is assumed to be uniform in the whole formation region of \MgII\
lines, which is certainly a crude approximation. 
The central line intensities are comparable to the observed ones at about 17:46~UT. 
The line cores are not
much reversed, consistently with the observations. We did not perform a convolution
with the IRIS instrumental profile with a width of 52 m\AA; doing so, one
could get the emission peaks slightly depressed towards the observed profile
\citep[see \eg][]{Heinzel2015}. 
The synthetic line profiles are significantly broadened by the microturbulence only
in the core, not in the wings. This means that, at least with this schematic model, 
the observed bright wings are due to an enhanced line source
function rather than due to an extra broadening.
This can be seen from the 
formation heights of the \MgII\ resonance lines. In Figure~\ref{fig:cf}
we show the contribution functions of the \MgII\ {\it k} line where broad 
line wings corresponding to the F2 atmosphere are formed over an extended region 
of about 1000 km. In the quiet Sun, \eg\ 
VAL C atmosphere \citep{Vernazza1981} which is shown for a comparison, the line wings are formed much deeper
in the atmosphere and over a region of only some 200 km wide. Finally, we also tested
the influence of the coronal pressure on the line spectrum. Model F2 has 
an upper boundary gas pressure equal to about 95 dyn cm$^{-2}$. The test solution
without this pressure leads to a much lower central intensity of the {\it k} and {\it h} lines.
Therefore, the overlying hot loops seem to have an important influence on the
line-core emission. The enhanced pressure is related to the explosive 
evaporation process, described in the next paragraph.

\begin{figure} 
   \centerline{\hspace*{0.015\textwidth}
               \includegraphics[width=0.5\textwidth,clip=]{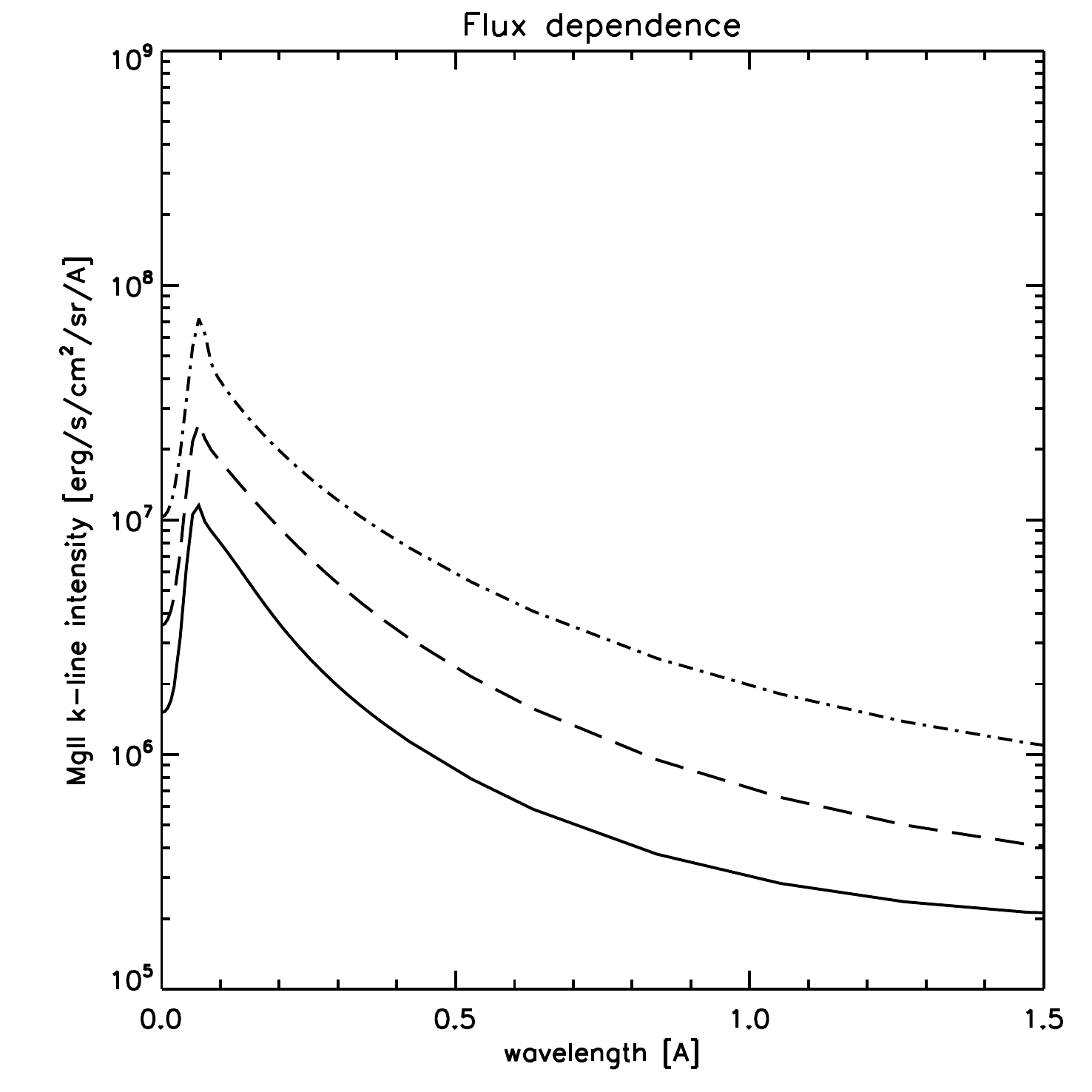}
               \includegraphics[width=0.5\textwidth,clip=]{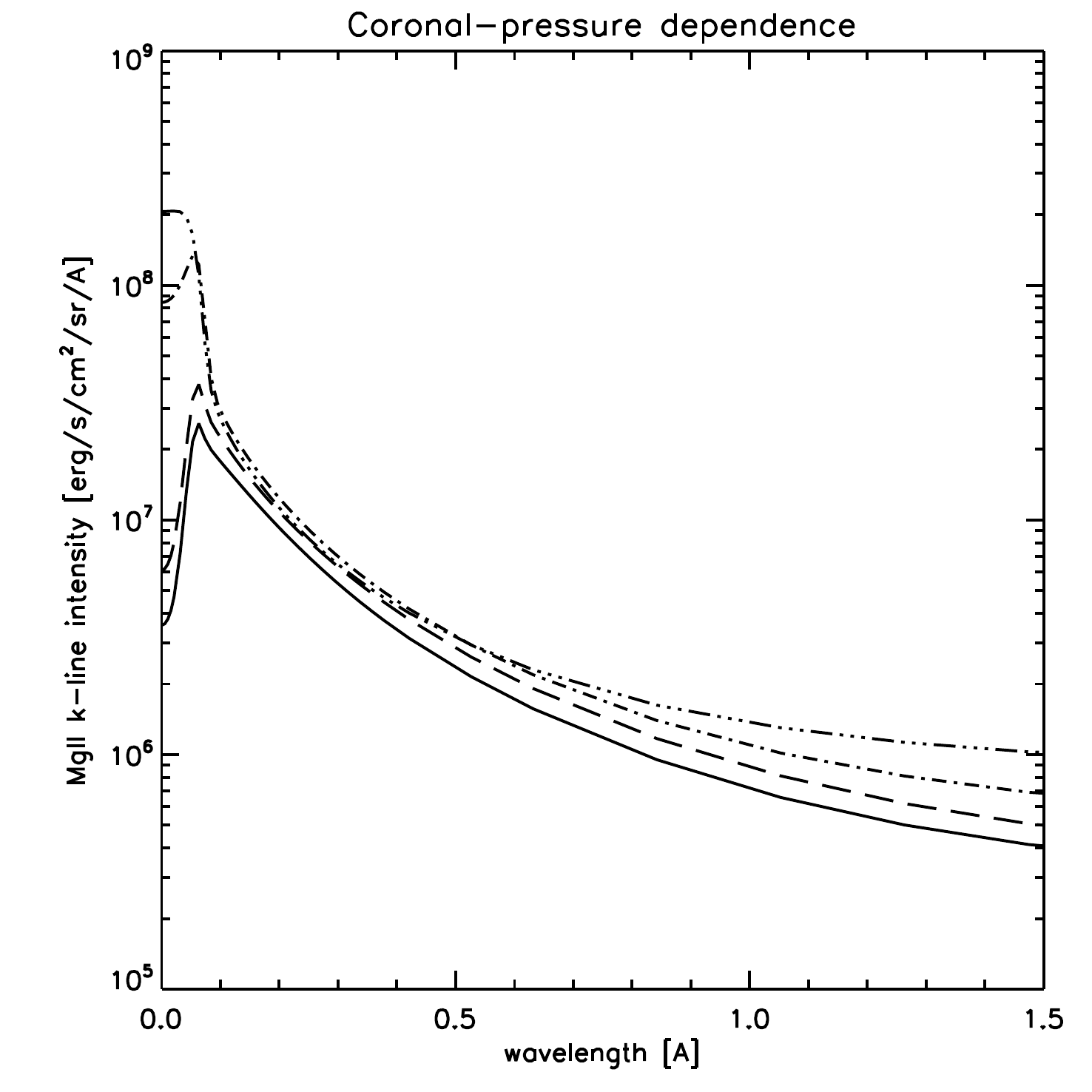}
              }
     \vspace{-0.44\textwidth}   
     \centerline{\Large \bf     
      \hspace{0.38\textwidth}  \color{black}{\small (a)}
      \hspace{0.42\textwidth}  \color{black}{\small (b)}
         \hfill}
   \vspace{0.43\textwidth}   
   \centerline{\hspace*{0.015\textwidth}
               \includegraphics[width=0.5\textwidth,clip=]{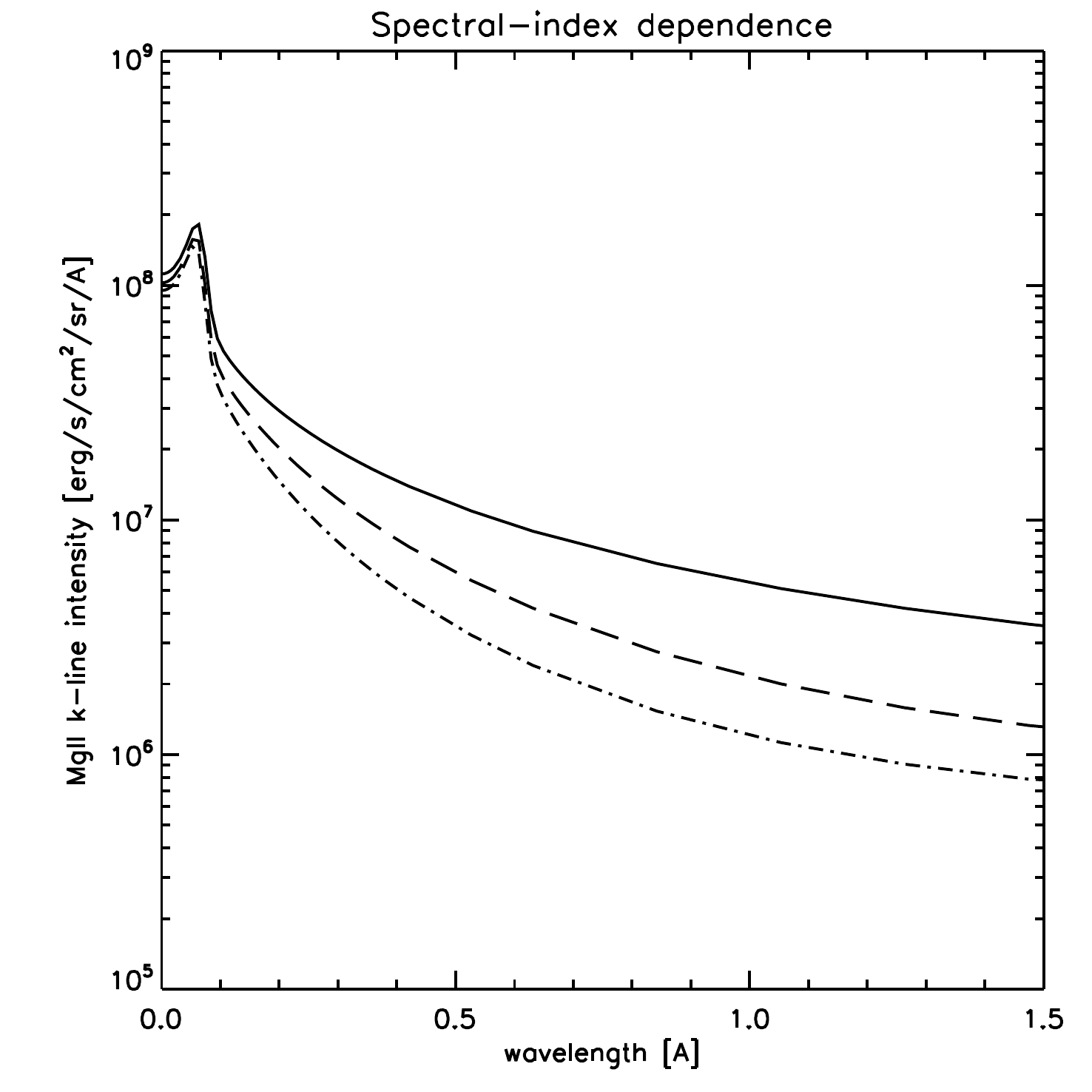}
              }
     \vspace{-0.44\textwidth}   
     \centerline{\Large \bf     
      \hspace{0.6\textwidth}  \color{black}{\small (c)}
         \hfill}
       \vspace{0.4\textwidth}    
\caption{\MgII\ {\it k} line profiles computed from RC models with different parameters. (a) $\delta$=5, 
coronal pressure 1 dyn cm$^{-2}$ and three electron-beam fluxes 10$^9$ (solid line), 10$^{10}$ (dashed line) 
and 10$^{11}$ erg s$^{-1}$ cm$^{-2}$ (dash-dotted line). (b) $\delta$=5, flux 10$^{10}$ erg s$^{-1}$ cm$^{-2}$ 
and four different values of the coronal pressure 1 (solid line), 10 (dashed line), 100 (dash-dotted line) and 1000 dyn cm$^{-2}$ 
(dash-three-dotted line). (c) the electron-beam flux 10$^{11}$ erg s$^{-1}$ cm$^{-2}$, coronal pressure 100 dyn cm$^{-2}$ 
and three spectral indexes of the electron beam $\delta$=3 (solid line), $\delta$=5 (dashed line) and
$\delta$=7 (dot-dashed line).}
   \label{fig:RC}
\end{figure}

To demonstrate the dependence of \MgII\ line intensities and line profile shapes on
various model parameters, we have also computed a grid of profiles based on models
by \citet{RC1983} (hereafter called RC models). 
Although these models have been published more than three decades ago, they still represent the only systematic grid of theoretical flare models.  
RC models are 1D static atmosphere models in stationary energy balance where the chromosphere is heated by electron beams.
The latter are parameterized by the total energy flux above a low-energy cutoff of 20 keV and a spectral index $\delta$. Another parameter is the coronal
pressure at the top of the model atmosphere which is supposed to be enhanced as a result of explosive evaporation.
Using RC models, we have computed synthetic \MgII\ {\it k} line profiles displayed in Figure~\ref{fig:RC}.
Figure~\ref{fig:RC}a shows the line-profile variations with increasing electron-beam flux, from 10$^9$ to 10$^{11}$ erg s$^{-1}$ cm$^{-2}$.
The spectral index is equal to five and the coronal pressure is 1 dyn cm$^{-2}$. All profiles are rather strongly reversed although for higher fluxes, the 
averaged line-core intensities are comparable to the observed ones. The wings are however too low. The central line reversal is significantly
reduced for higher coronal pressures, as shown in Figure~\ref{fig:RC}b. For the highest value of the pressure, the profile is non-reversed, but 
its central intensity is much higher compared to observations. Finally, Figure~\ref{fig:RC}c shows how the wing
intensities increase with decreasing $\delta$, while the line core remains practically unchanged. At $\delta$=3 the wing reaches the observed
intensities, which indicates a deep penetration of the beam into lower atmospheric layers, which are then more heated, contrary to models with a higher 
$\delta$ or the semi-empirical model F2. 
From this analysis we can conclude that a model with higher flux, between 10$^{10}$ and 10$^{11}$ erg s$^{-1}$ cm$^{-2}$, a spectral index 
around three, and and enhanced coronal pressure between 10 -- 100 dyn cm$^{-2}$ could roughly reproduce the observations. Such fluxes and spectral indexes are
consistent with those derived for this flare from detailed analysis of RHESSI spectra obtained at the same time and locations as
the \MgII\ spectra reported in this article \citep{Kleint2015b}.

\begin{figure} 
 \centerline{\includegraphics[width=1\textwidth,clip=]{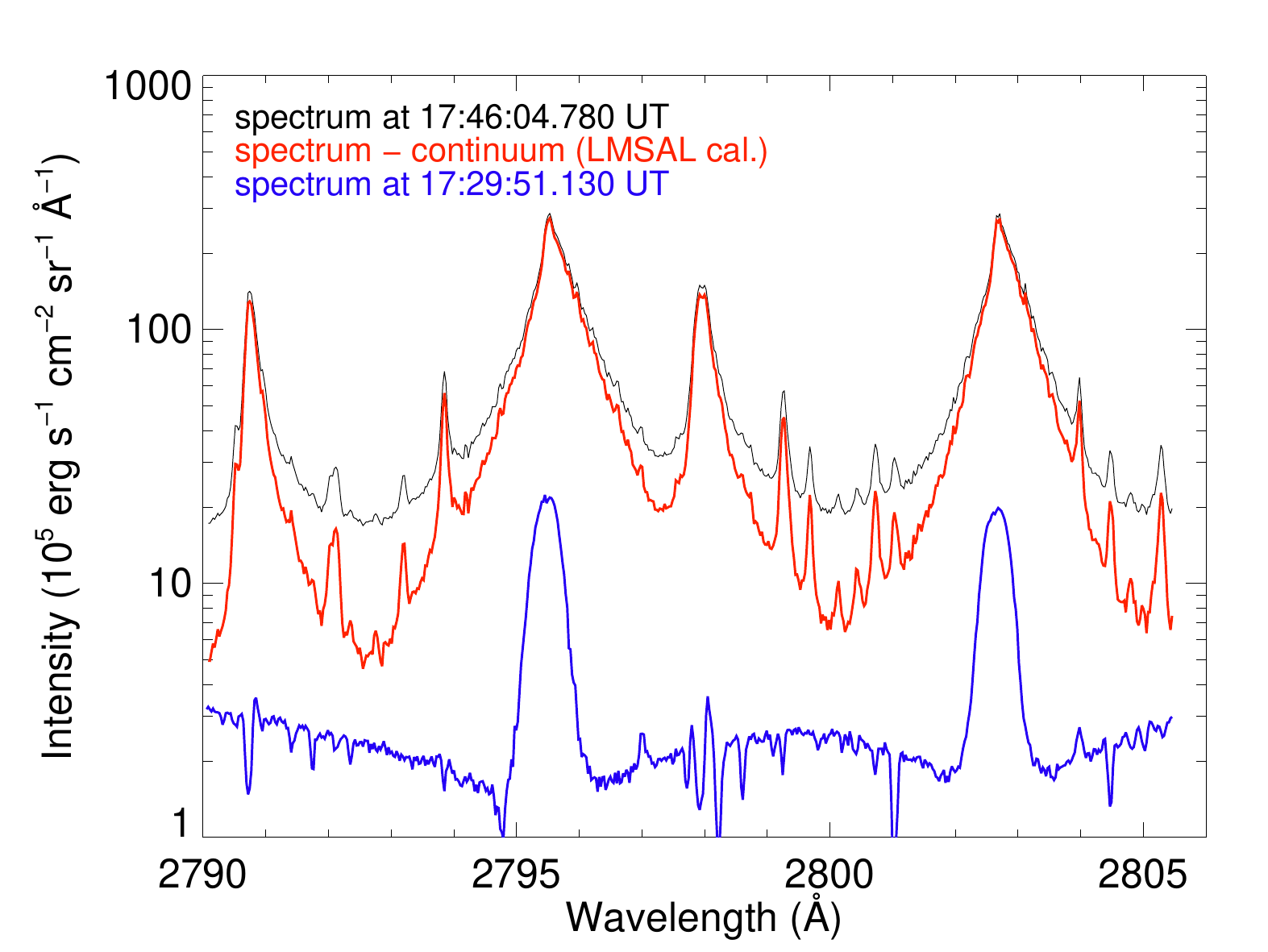}}
 \caption{Observed \MgII\ spectrum at 17:46:04.780~UT with (red) and without (black) the Balmer continuum
subtracted at pixel 447 slit {N$^0$} 4. 
The pre-flare spectrum at pixel 447 is also plotted in the figure with a blue continuous line, as a reference.}\label{fig:cont}
\end{figure}

There is still another aspect to be taken into account when one compares the
enhanced wings of the {\it k} and {\it h} lines. As found by \citet{Heinzel2014}, 
there seems to be a rather significant Balmer-continuum emission during the maximum of this
flare detected in the NUV channel of IRIS. Assuming that this emission is optically 
thin in the chromosphere, one can subtract it from the NUV spectrum by using the Lockheed Martin Solar and Astrophysics Laboratory (LMSAL) pre-flight calibration for the IRIS NUV spectra as in \citet{Heinzel2014}. This is
shown in Figure~\ref{fig:cont}, where the Balmer continuum is subtracted from the spectrum
at 17:46:04.78~UT (depending on the radiometric calibration used \citep{Heinzel2014} 
one gets the continuum levels differing by about a factor of three).
The effect of the Balmer continuum is marginal in the line core and inner wings 
where the line intensity is high, but it is significant
in the extended wings. Since our preliminary synthesis of the \MgII\ lines does not account
for an enhanced Balmer continuum of hydrogen, we should compare the synthetic profiles
from Figure~\ref{fig:F2} with the reduced ones in this Figure~\ref{fig:cont}. This naturally leads to a somewhat better
agreement. Note that when calculating the integrated intensities for the subordinate line or metallic line, we subtract
the continuum intensities at the wings near the center of the subordinate line or the metallic line before integration. So the integrated intensities in Figure~\ref{fig:sub-int} and Figure~\ref{fig:metal-int} are not affected by the Balmer continuum. %

\section{Conclusions}

We study the 2D spatial and temporal evolution of flare ribbons in
\MgII\ lines in detail.
The observations of the \MgII\ spectrum along all eight slits can be summarized as follows. (1) The intensities of 
the \MgII\ lines at the footpoints increase during the impulsive phase of the flare and peak at the maximum of the HXR emission. 
(2) The $k$ to $h$ ratio of the integrated intensities is 1.1 $\pm$ 0.068 at the footpoints and does not vary during the flare. 
(3) There are red asymmetries at most footpoints located in the southern ribbon after the peak. 
(4) The \MgII\ {\it h} and {\it k} lines are non-reversed at some footpoints and reversed at others 
at the flare peak. 
Some of the \MgII\ footpoints with reversed profiles are correlated with HXR sources. 
After the peak, the  \MgII\ line profiles at most footpoints become non-reversed. 
(5) The subordinate lines and metallic lines at the footpoints rise from absorption to emission during the flare. 
(6) The \MgII\ lines become broader during the flare. 

We have compared the observed line intensities with synthetic spectra
computed from the semi-empirical model F2 and found several similarities.
The temperature structure of the F2 model, which can be
regarded as a `snapshot' of a more realistic model evolution, reproduces some features
of the observed profiles, but does not explain the broad line wings well, which
will require stronger heating deeper in the chromosphere. 
Since the original F2 model was
constructed without taking into account the non-thermal rates in hydrogen, we also
neglected this aspect. For the \MgII\ lines, the non-thermal processes seem to be of
a secondary importance as we found from several tests. 
Another class of models is represented by a grid of theoretical flare models constructed by RC. For these
models we have synthesized the \MgII\ {\it k} profiles and shown that for some specific model parameters
they roughly fit our observations. However, all used models are static and thus they do not reproduce the
line asymmetries observed in some pixels. These mostly red asymmetries are usually associated with the 
downward moving chromospheric plasma. The velocity field affects the line profiles (asymmetry), but
also significantly influences the line intensities \textit{via} the statistical equilibrium for the \MgII\ atom.
Without any systematic modeling, it is rather difficult to interpret the observation that 
the \MgII\ line profiles are grouped into reversed (red in Figure~\ref{fig:rhessi}) and non-reversed (blue
in Figure~\ref{fig:rhessi}). However, based on a limited grid of RC models, we may speculate that reversed
profiles will appear at sites of strong heating and moderate coronal pressures where
the chromospheric condensation (downflows) is formed due to the electron-beam impact, leading to the
red asymmetries which are observed. On the other hand, unreversed and more symmetrical
profiles might be related to higher coronal pressures (the loops are
filled with the evaporated plasma) and motions inside the ribbons are very small.
This might be consistent with Figure~\ref{fig:rhessi} where the outer edges of the southern ribbon
may represent the footpoints of hot loops in which the electron beams propagate, while
the inner edges are places where cooler and denser loops are rooted. From the modeling we
found that the line cores are very sensitive to gas pressure in overlying coronal loops.
In a future study we
plan to perform more realistic radiative-hydrodynamical simulations of the temporal
evolution of flare spectra using the code {\em Flarix} developed at the Ond\v{r}ejov 
observatory \citep{Varady2010}. 
Electron-beam spectra derived
from simultaneous RHESSI observations \citep[see][]{Kleint2015b} will be used. These simulations
will also allow us to study the velocity flows, which cause the line asymmetries.

%

%

%
\begin{acks}

The research leading to these results has received funding from the European 
Community´s Seventh Framework Programme (FP7/2007--2013) under grant agreement 
N$^0$ 606862 (F-CHROMA) 
and from ASI ASCR project RVO:67985815. L.K. was supported by a Marie-Curie 
Fellowship. {\em IRIS} is a NASA small explorer mission developed and operated
by LMSAL with mission operation executed at the NASA Ames Research Center
and major contributions to downlink communications funded by the Norwegian
Space Center (NSC, Norway) through an ESA PRODEX contract.\\

\textbf{Disclosure of Potential Conflicts of Interest}\\

The authors declare that they have no conflicts of interest.

\end{acks}

%
%
%

%
%
%

\end{article} 
\end{document}